\documentclass[10pt, twocolumn,prd, showpacs, floatfix, 
preprintnumbers, nofootinbib, superscriptaddress]{revtex4}
\usepackage{epsfig}
\usepackage{setspace}
\usepackage{natbib}
\usepackage{booktabs, tabularx} 
\usepackage[normalem]{ulem}
 
\usepackage{amsmath,bm,amssymb}
\usepackage{color}
\definecolor{orange}{rgb}{1,0.5,0}

\newcommand{\code}[1]{\texttt{#1}}

\def\lsim{\ \raisebox{-.4ex}{\rlap{$\sim$}} \raisebox{.4ex}{$<$}\ }

\newcommand{\be}{\begin{equation}}  
\newcommand{\ee}{\end{equation}}
\newcommand{\bea}{\begin{eqnarray}}  
\newcommand{\eea}{\end{eqnarray}}

\newcommand{\dma}{{\Delta m_{\rm atm}^2}}

\newcommand{\nub}{{\bar{\nu}}}
\newcommand{\nue}{{\nu_e}}
\newcommand{\nui}{{\nu_i}}
\newcommand{\num}{{\nu_{\mu}}}
\newcommand{\nut}{{\nu_{\tau}}}
\newcommand{\nux}{{\nu_x}}

\newcommand{\nueb}{{\bar{\nu}_e}}

\newcommand{\nuxb}{{\bar{\nu}_x}}

\newcommand{\ie}{\emph{i.e.,}}

\newcommand{\etal}{\emph{et al.}}

\def\apjl{ApJ}                
\def\aap{A\&A}                
\def\physrep{Phys.~Rep.}   
\def\apjs{ApJS}                
\def\mnras{Mon. Not. Roy. Astron. Soc.}
\def\araa{Ann. Rev. Astron. Astrophys.}

\begin{document}  
  
\title{\mbox{\hspace{-0.75cm}The Role of Collective Neutrino Flavor Oscillations in Core-Collapse Supernova Shock Revival}}

\author{Basudeb Dasgupta} \affiliation{CCAPP, The Ohio State
  University, 191 W. Woodruff Avenue, Columbus, OH 43210, USA}

\author{Evan P. O'Connor}
\affiliation{\mbox{TAPIR, California Institute of Technology, MC 350-17, 1200 E. California Blvd., Pasadena, CA 91125, USA}}

\author{Christian D. Ott}
\affiliation{\mbox{TAPIR, California Institute of Technology, MC 350-17, 1200 E. California Blvd., Pasadena, CA 91125, USA}}

\date{15 September, 2011}  
\preprint{}   
  
\begin{abstract}   
  We explore the effects of collective neutrino flavor oscillations
  due to neutrino-neutrino interactions on the neutrino heating behind
  a stalled core-collapse supernova shock.  We carry out axisymmetric
  (2D) radiation-hydrodynamic core-collapse supernova simulations,
  tracking the first 400~ms of the post-core-bounce evolution in
  $11.2$-$M_\odot$ and $15$-$M_\odot$ progenitor stars. Using inputs
  from these 2D simulations, we perform neutrino flavor oscillation
  calculations in multi-energy single-angle and multi-angle
  single-energy approximations. Our results show that flavor
  conversions do not set in until close to or outside the stalled
  shock, enhancing heating by not more than a few percent in the most
  optimistic case. Consequently, we conclude that the postbounce
  pre-explosion dynamics of standard core-collapse supernovae remains
  unaffected by neutrino oscillations. Multi-angle effects in regions
  of high electron density can further inhibit collective
  oscillations, strengthening our conclusion.
\end{abstract}   

\maketitle

\section{Introduction}         \label{intro}  
Stars with a zero-age main-sequence (ZAMS) mass
greater than $\sim$(8$-$10)-$M_\odot$ 
undergo core collapse at the end of their
lives.  When the inner core reaches nuclear density, the stiffening of
the nuclear equation of state (EOS) induces core bounce, launching a
strong shock wave that slams into the still infalling outer
core. Created at a mass coordinate of $\sim$$0.5$-$M_\odot$ and endowed
with the kinetic energy from infall, the shock initially moves out
rapidly in mass and radius, but dissociation of heavy infalling nuclei
and neutrino losses from neutronization and thermal processes in the
hot region behind the shock sap its energy. As a result, the shock soon
succumbs to the ram pressure of the outer core and stalls at a radius
of~$\sim$$(100 - 200)\,\mathrm{km}$.

Finding the mechanism that robustly revives the stalled shock to blow
up a massive star in a core-collapse supernova has been the primary
objective of core-collapse supernova theory for decades, 
but so far success has been limited. The \emph{neutrino
  mechanism}~\cite{bethewilson:85,bethe:90}, based on the net
deposition of neutrino energy by charged-current absorption in the
semi-transparent \emph{gain} region below the shock, has been shown by
radiation-hydrodynamics simulations to work in its purest,
spherically-symmetric (1D) form only in the lowest-mass massive stars
with oxygen-neon cores~\cite{kitaura:06,huedepohl:10,burrows:07c} and
the most accurate 1D simulations fail to explode more massive
stars~\cite{liebendoerfer:05,thompson:03}. In axisymmetry (2D),
neutrino-driven convection and the standing-accretion-shock
instability (SASI, \emph{e.g.}, \cite{blondin:03,scheck:08,foglizzo:07})
increase the efficacy of the neutrino mechanism (by enhancing heating
\cite{murphy:08,buras:06a,marek:09} or reducing cooling by
neutrinos~\cite{pejcha:11}).  Neutrino-driven explosions in 2D
simulations have been reported~\cite{marek:09,bruenn:09,suwa:10},
though only in models using the softest variant of the Lattimer-Swesty
EOS~\cite{lseos:91}, which is disfavored by the recent discovery of a
$2$-$M_\odot$ neutron star \cite{demorest:10},
but leads to a compact
protoneutron star (PNS) and a hard neutrino
spectrum~\cite{marek:09b,marek:09}, favorable for heating due to the
$\epsilon_\nu^2$ dependence of the neutrino absorption~cross~section.

Other multi-dimensional phenomena have led to proposals of
alternatives to the neutrino mechanism: Rapid rotation in combination
with rotational magnetic field amplification can lead to
magnetorotational explosions with bipolar morphology
\cite{leblanc:70,bisno:70,burrows:07b,dessart:08a}. But rapid core
rotation may be present only in a small fraction of all massive stars
\cite{heger:05,ott:06spin}, ruling out this \emph{magnetorotational
  mechanism} for the garden-variety core-collapse supernova.
Burrows~\emph{et~al.}~\cite{burrows:06,burrows:07a} proposed an
\emph{acoustic mechanism} in which non-linear PNS oscillations, driven
by turbulence and accretion downstreams, emit sound waves into the
region behind the shock that steepen to secondary shocks, injecting
heat and reviving the stalled shock. The acoustic mechanism leads to
perhaps too late, too weak explosions and bleeding of PNS oscillation
power to numerically unresolved daughter modes may diminish its
relevance in nature~\cite{weinberg:08}.

Exploratory 3D simulations \cite{fryerwarren:02,nordhaus:10} are
suggestive of the possibility that the additional degree of freedom
over 2D and the physical nature of turbulence in 3D could render the
neutrino mechanism robust. But full 3D neutrino
radiation-hydrodynamics simulations must be awaited before 3D can be
declared the missing piece in the core-collapse supernova puzzle.

The marginality of the various proposed mechanisms combined with
nature's robust ability to produce explosions in massive stars up to
at least $\sim$$20$-$M_\odot$ \cite{smartt:09b,oconnor:11} makes one
wonder: \emph{Is there important physics missing from core-collapse
  supernova models?} With the current standard set of physics included
in models, we might, for example, miss an early hadron-quark phase
transition in the PNS, leading to a second collapse and bounce and a
second shock wave that revives the first. This possibility was proposed
by \cite{sagert:09,fischer:11}, but requires a soft hadronic EOS that
has now been ruled out.

In this article, we consider new physics that has not previously been
included in the core-collapse supernova problem: self-induced
collective neutrino flavor oscillations. Neutrino oscillations have
long been known to occur in vacuum (\emph{e.g.}, \cite{Pontecorvo:1967fh})
and in matter, mediated by neutrino-electron scattering (via the
Mikheyev-Smirnov-Wolfenstein (MSW) effect,
\cite{Wolfenstein:1977ue,Mikheev:1986gs}). Less appreciated, until
recently, has been the possibility of oscillations occurring due to
neutrino-neutrino forward scattering in regions of high neutrino
number density. This self-induced oscillation process was first
discussed in~\cite{Pantaleone:1992eq} and then explored in a series of
papers~\cite{Samuel:1993uw, Kostelecky:1994dt, Pastor:2001iu,
  Wong:2002fa, Abazajian:2002qx,
  Friedland:2003dv,Sawyer:2005jk,Fuller:2005ae}. Thereafter, following
the simulations of Duan \etal~\cite{Duan:2005cp,Duan:2006an}, it has
received much attention recently as a process occurring in the
core-collapse supernova environment and leading to flavor conversions
of neutrinos and antineutrinos of almost all energies (\emph{e.g.},
\cite{Hannestad:2006nj,Duan:2007mv, Raffelt:2007cb,
  EstebanPretel:2007ec,Fogli:2007bk, Dasgupta:2007ws, Duan:2008za,
  Dasgupta:2008cd, Dasgupta:2008my, EstebanPretel:2008ni, Gava:2009pj,
  Dasgupta:2009mg, Dasgupta:2010ae, Raffelt:2010za, Cherry:2010yc,
  Duan:2010bf, Mirizzi:2010uz, Galais:2011jh, Raffelt:2011yb,
  Pehlivan:2011hp}. Also see~the~review~\cite{Duan:2010bg}).

The most intriguing result of these so-called ``collective
oscillations'' (flavor conversions take place collectively over all
energies) is an almost complete exchange of electron neutrino
($\nu_e$) and antineutrino ($\bar{\nu}_e$) spectra with the spectra of
the heavy-lepton neutrinos and antineutrinos $\nu_x \in \{\nu_\mu,
\nu_\tau, \bar{\nu}_\mu,\bar{\nu}_\tau\}$. The $\nu_x$, due to the
absence of muons and taus, do not interact via charged-current
processes in core-collapse supernovae. They are created by thermal
processes in the PNS core and decouple from matter at smaller radii
and higher temperatures than $\nu_e$ and $\bar{\nu}_e$.  Hence, the
initial $\nu_x$ spectra are much harder. Due to the $\epsilon_\nu^2$
dependence of the charged-current absorption cross section, a swap of
$\nu_x$ and $\nu_e$/$\bar{\nu}_e$ spectra would dramatically enhance
neutrino heating and may be the crucial ingredient missing in
core-collapse supernova models, provided the swap occurs at
sufficiently small radii in the region behind the shock to boost net
heating. This is indeed the recent result obtained by Suwa~\emph{et
  al.}~\cite{suwa:11}, where a neutrino conversion radii was assumed
at a radius of $100\,\mathrm{km}$.

Fuller~\emph{et~al.}~\cite{Fuller:1992} were the first to propose
increased heating due to neutrino oscillations, but their MSW resonance 
based oscillation mechanism required a large neutrino mass of $\sim
(10-100)\,\mathrm{eV}$ for one of the active neutrinos.
Akhmedov~\emph{et~al.}~\cite{Akhmedov:1996ec} put forth a similar
proposal, but both are now ruled out by stringent constraints on
neutrino masses~(\emph{e.g.}, \cite{Otten:2008zz}).  

Self-induced collective neutrino oscillations, on the other hand, do
not require large neutrino masses, are a rather straightforward
consequence of the Standard Model of particle physics, and may, in
principle, occur in both the inverted and the normal neutrino mass
hierarchy~\cite{Dasgupta:2009mg}. In this work, we study their
relevance for the core-collapse supernova mechanism by calculating
approximate analytic and detailed numerical estimates for the radii at
which collective oscillations set in and could influence neutrino
heating. We base these calculations on neutrino radiation fields from
2D neutrino radiation-hydrodynamic simulations of the postbounce
core-collapse supernova evolution in $11.2$-$M_\odot$ and
\mbox{$15$-$M_\odot$} progenitor stars, representative of the
progenitors of standard Type-II supernovae.  

We find that the calculated oscillation radii, while reaching average
shock radii, due not penetrate deeply into the heating region. Large
shock excursions due to the SASI reach and surpass the radius at which
oscillations set in, but the region in which the vast majority of net
heating occurs remains always at least $\sim$$100\,\mathrm{km}$ below
the oscillation radius, even in the low-mass \mbox{$11.2$-$M_\odot$}
progenitor.  Recent results of
Chakraborty~\emph{et~al.}~\mbox{\cite{Chakraborty:2011nf,
    Chakraborty:2011gd}}, obtained on the basis of 1D simulations,
show the suppression of collective oscillations by very high electron
number density in dense matter. In our 2D models, we find that such a
suppression may be significantly weaker than reported by
Chakraborty~\emph{et al.}. Nonetheless, the flavor conversion still
happens too far out in the supernova, and we conclude that collective
neutrino oscillations do not have a significant effect on the
explosion mechanism of core-collapse supernovae in progenitors in and
above the explored~mass~range.

The structure of this article is as follows.  In Section
\ref{sec:revival}, we review neutrino heating in core-collapse
supernovae and in Section \ref{collective}, we introduce collective
neutrino oscillations and present an approximate analytic prescription
that can be used to determine the radius at which neutrinos will begin
to collectively oscillate. In Section \ref{sec:numerical}, we go on to
discuss our 2D radiation-hydrodynamic postbounce core-collapse
supernova simulations and contrast the evolutions of their
characteristic radii with the analytic estimates and detailed
numerical results for the oscillation radius. This allows us to
ascertain the importance of collective neutrino oscillations for shock
revival. In Section \ref{sec:conclusions}, we summarize our~results~and~conclude.

\section{Neutrino Heating in Core-Collapse Supernovae}  
\label{sec:revival}  

To elucidate the basics of the neutrino mechanism, we make a number of
simplifying assumptions, which we lay out in the following. We assume
a spherically symmetric mass distribution and expect neutrinos to
stream freely outside their energy-averaged neutrinospheres. We define
the neutrinosphere radii for each neutrino species via a Rosseland
mean neutrino optical depth,
\begin{equation}
  \tau_{\mathrm{RM}, \nu_i}(r) =
  \int_r^\infty \left(\frac{\int_0^\infty J_{\nu_i}/\kappa_{\nu_i}\,
      d\epsilon}{\int_0^\infty J_{\nu_i}\, d\epsilon}\right)^{-1} dr^\prime \,.\label{eq:RM}
\end{equation}
and set the neutrinosphere radius $R_{\nu_i} = R(\tau_{RM,\nu_i} =
2/3)$. In this expression for the energy-averaged optical depth,
$\kappa_{\nu_i}$ is the sum of the absorption and scattering
opacities, and $J_{\nu_i}$ is the $\nu_i$ energy density.  

The energy-dependent optical depth is a quadratic function of the
neutrino energy, the energy-dependent neutrinospheres move outward
with neutrino energy (cf. Fig.\,13 of \cite{dessart:06pns}). The
energy-averaged variant of the optical depth that we use produces a
neutrinosphere that matches rather well the radius of the
neutrinosphere of the average neutrino energy for each species. In the
following, we will consider only the $\nu_e$ neutrinosphere radius
$R_{\nu_e}$, since $\nu_e$ decouple from matter furthest out, followed
first by $\bar{\nu}_e$, then by $\nu_x$. The $\nux$ are not
involved in charged-current interactions, hence have the largest mean
free path. The $\nueb$ interact with the less abundant
proton and, hence, also have a greater mean free path than the
$\nu_e$. This decoupling hierarchy is also present in the mean and
mean-squared neutrino energies, giving $\langle \epsilon_{\nu_x}
\rangle > \langle \epsilon_{\bar{\nu}_e} \rangle > \langle
\epsilon_{\nu_e} \rangle$ and $\langle \epsilon^2_{\nu_x} \rangle >
\langle \epsilon^2_{\bar{\nu}_e} \rangle > \langle \epsilon^2_{\nu_e}
\rangle$, reflecting the fact that the least interacting neutrino
decouples at the smallest radius and the highest matter
temperature. Typical values for the mean neutrino energies, $\langle
\epsilon_{\nu_i} \rangle$, are $\sim$$(10-15)$~MeV for $\nue$, and
$\nueb$ and $\sim$$(15-20)$~MeV for $\nux$~(see, \emph{e.g.},
\cite{ott:08,marek:09b,brandt:11}).  In the very early postbounce
phase of a core-collapse supernova, $R_{\nu_e}$ is typically around
\mbox{$\sim$$(70-80)\,$km}, reaching \mbox{$\sim$$(30-40)\,$km} within \mbox{$(200-300)\,$ms}
of core bounce.

We make the assumption that beyond our nominal neutrinosphere
$R_{\nu_e}$, the radiation fields of all neutrinos are freely
streaming with a luminosity $\mathcal{L}_{\nu_i}$, an average energy
of $\langle \epsilon_{\nu_i} \rangle$, a mean squared energy of
$\langle \epsilon_{\nu_i}^2\rangle$, and a total \emph{number
  luminosity} of ${\cal N}_\nui = 4\pi R_{\nu_e}^2 \Phi_\nui$, where
  $\Phi_{\nu_i}$ is the neutrino number flux of species $i$ at radius
  $R_{\nu_e}$. We normalize the spectral neutrino distribution
  function $d\Phi_{\nu_i} / d\epsilon$ at the $\nu_e$ neutrinosphere
  according to $4\pi R_{\nu_e}^2 \int_0^\infty d\epsilon\,
  d\Phi_{\nu_i} / d\epsilon = 4\pi R_{\nu_e}^2 \Phi_{\nu_i}$.  The
  neutrino energy distribution function is then simply $\epsilon
  d\Phi_{\nu_i} / d\epsilon$.

Outside of $R_{\nu_e}$, conditions arise where absorption of $\nue$ and
$\nueb$ via charged-current interactions with neutrons and protons
inject more energy into the matter than is lost due to thermal
emission and electron and positron capture on neutrons and protons.
This is the case in the \emph{gain region} \cite{janka:01}, which, in
our simplified 1D picture, extends from the gain radius
$R_\mathrm{g}$, where neutrino heating balances cooling, to the shock
radius $R_\mathrm{s}$. Typical values of $R_\mathrm{g}$ and
$R_\mathrm{s}$ during the accretion phase of core-collapse supernovae
are $\sim$$100\,$km, and $\sim$$200\,$km, respectively~\mbox{\cite{buras:06a,buras:06b,
marek:09,ott:08}}.

In the gain region, the heating due to the charged-current neutrino-matter
interactions is given by the absorption cross section
${\sigma_{\nu_i}}(\epsilon_{\nu_i})$ convolved with the spectral 
energy fluxes of both $\nue$s and $\nueb$s incident on
the gain region from below, 
\begin{equation}
 {\cal H}  =  \sum_{\nue,\nueb}\int_{r_\mathrm{g}}^{r_\mathrm{s}} dr\,4\pi
 r^2\,n_{i}\int_{0}^{\infty} d\epsilon \,
 {\sigma_{\nu_i}}(\epsilon)\,\frac{\epsilon d\Phi_{\nu_i}}{d\epsilon}\,,
\label{eq:heating_one}
\end{equation}
where $n_{i}$ is the local number density of nucleons relevant for
interactions with neutrino species $i$. For Eq.\,(\ref{eq:heating_one}),
we approximate  $\sigma_{\nu_i}(\epsilon) = \sigma_0
(1+3g_\mathrm{A}^2) /(4 m_e^2 c^4) \times \epsilon^2 = \hat{\sigma}
\epsilon^2$, where $\sigma_0 \sim$$1.76\times10^{-44}\, \mathrm{cm}^2$
is the reference weak interaction cross section, and $g_\mathrm{A}$ is
the axial-vector coupling constant. Note that the units of
$\hat{\sigma}$ are cm$^2$/MeV$^2$. This approximation of the neutrino
absorption cross section, in combination with our free-streaming
assumption, leads to a gross heating rate,
\begin{eqnarray}
 {\cal H}& \sim & \sum_{\nu_i} \hat{\sigma}\langle \epsilon^2_{\nu_i}\rangle
 {\cal L}_{\nu_i} \int_{R_\mathrm{g}}^{R_\mathrm{s}} dr\,n_{i} \nonumber\\ 
 & \sim & \hat{\sigma} \left [ \langle
   \epsilon^2_{\nu_e}\rangle {\cal L}_{\nu_e}c_N + \langle
   \epsilon^2_{\bar{\nu}_e}\rangle {\cal L}_{\bar{\nu}_e}c_P \right]\,\,,
\label{eq:heatrate}
\end{eqnarray} 
where $c_N$ and $c_P$ are the target nucleon column densities (in
\#/cm$^2$) for $\nu_e$ and $\bar{\nu}_e$, respectively. Note that
${\cal L}_{\nu_i}$ and~$\langle \epsilon^2_{\nu_i}\rangle $ will
typically vary somewhat across the gain region. We neglect this
variation for simplicity. Eq.\,(\ref{eq:heatrate}) gives the
integrated gross heating rate, $\mathcal{H}$, but it is important to
note that the majority of the heating occurs very near the gain radius
due to the strong dependence of the rest-mass density on radius, $\rho
\propto\ r^{-3}$ (as pointed out by \cite{janka:01} and seen in
simulations. See, \emph{e.g.}, Fig.\,16 of \cite{ott:08}, and
Fig.\,\ref{fig:s11.2_and_s15-MA} of this work). Also, while neutrino
heating dominates over cooling in the gain region, the latter is still
significant and must not be neglected. The net heating rate (heating
minus cooling) is estimated by Janka \cite{janka:01} to be
\mbox{${\cal H}_\mathrm{net} = {\cal H} - {\cal C} \approx {\cal
    H}/2$}, which is in rough agreement with what
one~finds~in~simulations.

The main take-away message from this section is that the heating rate
${\cal H}$ depends on the $\nue$ and $\nueb$ spectral fluxes as 
\begin{equation}
{\cal H}\propto\sum_{\nu_i=\nue,\nueb}{\cal L}_{\nu_i}\langle
\epsilon^2_{\nu_i}\rangle\,.  
\end{equation}
Our aim in this study is to explore if flavor conversions due to
collective neutrino oscillations can increase the quantity ${\cal
  L}_{\nue,\,\nueb}\langle \epsilon^2_{\nue,\,\nueb}\rangle$ to boost
the heating mechanism.  There are two aspects to this question -- (i)
What is the typical degree of flavor conversion?  (ii) Does this
flavor conversion take place at sufficiently small radii to have a
significant effect on the total neutrino heating? To address these
questions, we discuss collective oscillations in the next section.

\vspace{-0.1cm}
\section{Collective Neutrino Flavor Conversion}         \label{collective}  
\subsection{Equations of Motion} 

Neutrinos with masses $m_1,\,m_2,\,m_3$ are related to three flavor
states $\nue,\, \num,\,\nut$. These flavor states oscillate from one
to another as a function of time, depending on the mass-square
differences $\Delta m_{ij}^2=m_j^2-m_i^2$ and mixing angles
$\theta_{ij}$, where the indices run over~($1,\,2,\,3$). For an
introduction to neutrino oscillation physics, see \emph{e.g.},~Chapter 3 of
~\cite{Strumia:2006db}. As is usual in particle physics, we will use
units in which the speed of light $c$ and Planck's constant $\hbar$
are equal to 1.

In the core-collapse supernova context, a two neutrino flavor
approximation is often appropriate because $\Delta m^2_{21}\ll|\Delta
m^2_{31}|$ and $\theta_{13}\ll1$.  In that case, the oscillations are
primarily between $\nue$ and the linear combination
$(\num-\nut)/\sqrt{2}$, while the other linear combination is
decoupled from the system. Our numerical results have been
  obtained with a full three-flavor code.  The
$\nue\leftrightarrow\nux$ and $\nueb\leftrightarrow\nux$ oscillations
are governed by $\dma\approx\Delta m^2_{31}$ (the subscript ``atm'' is
used, because this quantity determines the oscillations of neutrinos
created in the atmosphere) and the mixing angle $\theta_{13}$. We take
the absolute value of $\dma$ to be $2.6\times10^{-3}$~eV$^2$, close to
the best fit of experimental data~\cite{Schwetz:2011qt}, while the
sign is not known.  We will use a benchmark value of
$\theta_{13}=0.001$, consistent with the upper limit
$\sin^2\theta_{13}<0.035$~\cite{Schwetz:2011qt}.

Neutrino oscillations can alter the flux differences
$d\Phi_{\nue}/d\epsilon-d\Phi_{\nux}/d\epsilon$ and
$d\Phi_{\nueb}/d\epsilon-d\Phi_{\nux}/d\epsilon$ at a given energy.
We will represent the flux differences at each energy $\epsilon$ by a
polarization vector ${\bf P}$ and $\overline{\bf P}$ in a
three-dimensional flavor space, as in~\cite{Hannestad:2006nj}. At the
neutrinosphere, all neutrinos are emitted as flavor states. Thus the
initial polarization vectors are aligned with the \mbox{${z}$-direction} 
\bea
{\bf
  P}(\epsilon)&=&\frac{d\Phi_{\nue}/d\epsilon-d\Phi_{\nux}/d\epsilon}{\Phi_{\nue}+\Phi_{\nueb}+4\Phi_{\nux}}\,\hat{z}\,,\label{Pvec}\\ \overline{{\bf
    P}}(\epsilon)&=&\frac{d\Phi_{\nueb}/d\epsilon-d\Phi_{\nuxb}/d\epsilon}{\Phi_{\nue}+\Phi_{\nueb}+4\Phi_{\nux}}\,\hat{z}\,,\label{Pvecb}
\eea where a vertically upward vector represents a $\nue$ (or $\nueb$)
excess and a downward vector shows a $\nux$ excess. When the fluxes
of $\nue$ (or $\nueb$) and $\nux$ are equal, the polarization vector
vanishes. Other directions represent states that are coherent
superpositions of the two pure states which will be generated by
flavor oscillations.

In the treatment of neutrino oscillations it is important to consider
the angle $\vartheta_{R_{\nue}}$ at which neutrinos are emitted from
the neutrinosphere at $R_{\nue}$. We will therefore attach a label $u$
to polarization vectors signifying the direction of emission,
\ie~${\bf P}(\epsilon, u)$, where $u=\sin^2\vartheta_{R_{\nue}}$. Of
course, $\int_0^1 du {\bf P}(\epsilon, u)={\bf P}(\epsilon)$, and
similarly for antineutrinos.
  
In vacuum, the $\nue$ (or $\nueb$) oscillate to $\nux$ and back with a
frequency~\cite{Pontecorvo:1967fh} 
\bea
\omega(\epsilon)&=&\frac{|\dma|}{2\epsilon}\,,\\ 
&=&0.65~{\rm km}^{-1}\times\left[\frac{10\,{\rm
      MeV}}{\epsilon}\right]\,, 
\eea 
 solely under the action of neutrino masses and mixing. The vacuum Hamiltonian is
represented by a vector $\omega(\epsilon){\bf B}$ with ${\bf
  B}=\mp(\sin2\theta_{13}, 0, \cos2\theta_{13})$ where the minus sign
applies for normal neutrino mass hierarchy ($\dma>0$) and the plus
sign for inverted neutrino mass hierarchy ($\dma<0$).

In the core-collapse supernova environment there are additional
sources of flavor oscillation, \ie~weak interactions with the stellar
material, and weak interactions between the neutrinos themselves. In
the limit that neutrinos are free streaming, \ie~outside the
neutrinospheres, elastic scattering with electrons is the only
neutrino--matter process that is relevant. Of these,
the forward scattering amplitudes add coherently with the free
propagator to introduce a potential in the flavor evolution
Hamiltonian. This contribution due to the net local number density of
 electrons in the medium is known as the MSW
potential~\cite{Wolfenstein:1977ue,Mikheev:1986gs}, \bea
\lambda(r)&=&\sqrt{2}G_F
\left[n_e(r)-n_{e^+}(r)\right]\,,\\ &=&6.6\times10^{5}~{\rm km}^{-1}\times\left[\frac{n_e(r)-n_{e^+}(r)}{10^{33}/{\rm
      cm}^3}\right]\,, \eea where, in the second line, we
have used the value of the Fermi constant
$G_F=1.16\times10^{-5}$~GeV$^{-2}$. The matter-induced contribution to
the Hamiltonian is represented by a vector $\lambda(r){\bf L}$, where
${\bf L}=(0,\,0,\,1)$.

The above two contributions are well-known, and lead to the
traditional paradigm of core-collapse supernova neutrino oscillations
based on vacuum oscillations and matter-induced
oscillations~\cite{Dighe:1999bi}. However, near the neutrinosphere,
neutrino densities are very high, so in addition to ordinary neutrino
oscillations due to $\omega(\epsilon)$ and matter oscillations due to
$\lambda(r)$, one has appreciable forward scattering of neutrinos and
antineutrinos off each other~\cite{Pantaleone:1992eq}.  This
  leads to another potential, induced by all other neutrinos and
  antineutrinos, whose value at the neutrinosphere $R_{\nue}$ is given
  by~\cite{Pantaleone:1992eq,EstebanPretel:2007ec}
\bea
  \mu_{R_{\nue}}&=&\sqrt{2}G_F\,\Phi_{\nu,\nub}\,\\ &=&1.1\times10^6~{\rm
    km}^{-1}\phantom{\frac{1}{1}}\\ &\times&\left[\frac{(10\,{\rm
        km})^2}{R_{\nue}^2}\right]\sum_{\nu_i}\left[\frac{{\cal
        L}_{\nu_i}}{10^{52}\,{\rm erg/s}}\right]\left[\frac{10\,{\rm
        MeV}}{\langle\epsilon_{\nu_i}\rangle}\right]\,.\nonumber 
\eea

The exact quantity that appears here,
\ie~$\Phi_{\nu,\nub}=\Phi_{\nue}+\Phi_{\nueb}+4\Phi_{\nux}$, depends
on our chosen normalization for the polarization vectors --
Eqs.\,(\ref{Pvec}) and (\ref{Pvecb}) have the same quantity in the
denominator.  Altogether, the equations of motion
become~\cite{EstebanPretel:2007ec}

\begin{widetext}
\bea
\frac{d{\bf P}(\epsilon,u)}{dr}&=&+\frac{{\omega}(\epsilon){\bf
    B}\times{\bf  P}(\epsilon,u)}{v_r(u,r)}+\frac{{\lambda(r)}{\bf L}\times{\bf
    P}(\epsilon,u)}{v_r(u,r)}\label{eom-ma1}\\
    &&+\mu_{R_{\nue}}\frac{R_{\nue}^2}{r^2}\left[\left(\int_0^1du' \int_o^\infty d\epsilon'\frac{{\bf P}(\epsilon',u')-\overline{{\bf P}}(\epsilon',u')}{v_r(u',r)}\right)\times\frac{{\bf P}(\epsilon,u)}{v_r(u,r)} - \left(\int_0^{\infty}d\epsilon'\left({\bf P}(\epsilon')-\overline{\bf P}(\epsilon')\right)\right)\times{\bf P}(\epsilon,u)\right]\,,\nonumber\\ 
\frac{d\overline{\bf P}(\epsilon,u)}{dr}&=&-\frac{{\omega}(\epsilon){\bf
    B}\times\overline{\bf  P}(\epsilon,u)}{v_r(u,r)}+\frac{{\lambda(r)}{\bf L}\times\overline{\bf
    P}(\epsilon,u)}{v_r(u,r)}\label{eom-ma2}\\
    &&+\mu_{R_{\nue}}\frac{R_{\nue}^2}{r^2}\left[\left(\int_0^1du' \int_o^\infty d\epsilon'\frac{{\bf P}(\epsilon',u')-\overline{{\bf P}}(\epsilon',u')}{v_r(u',r)}\right)\times\frac{\overline{\bf P}(\epsilon,u)}{v_r(u,r)} - \left(\int_0^{\infty}d\epsilon'\left({\bf P}(\epsilon')-\overline{\bf P}(\epsilon')\right)\right)\times\overline{\bf P}(\epsilon,u)\right]\,.\nonumber    
\eea
\end{widetext}
Note that neutrinos emitted at angle $\vartheta_{R_{\nue}}$ have a radial velocity, 
\be
v_{r}(u,r)=\cos\vartheta(u,r)=\sqrt{1-u\frac{R_{\nue}^2}{r^2}}\,,
\ee
and their flavor evolution has been projected onto the radial direction.

To keep our discussion simple, we will often use an effective
spherically symmetric description proposed in
\cite{EstebanPretel:2007ec}, where all neutrinos are assumed to be
emitted at $45^\circ$ to the nominal neutrinosphere at
$R_{\nue}$. This is often referred to as the single-angle
  approximation. We will see that there is no flavor change close to
the neutrinosphere, thus this choice of a common neutrinosphere merely
acts as a boundary condition where we specify our initial states. In
this approximation, all neutrinos have a radial velocity
\be
v_{r}(r)=\sqrt{1-\frac{R_{\nue}^2}{2r^2}}\,.  
\ee 
The forward scattering amplitudes due to neutrinos and antineutrinos
scattering off each other leads to a collective potential, 

\be
  \mu(r)=\mu_{R_{\nue}}\times\left(\frac{R_{\nue}^2}{r^2}\right)\left(\frac{R_{\nue}^2/r^2}{2-R_{\nue}^2/r^2}\right)\,\,.
\label{eq:muofr}
\ee
The potential weakens as $1/r^4$ at large distances
because the fluxes dilute as $1/r^2$ and there is another
approximately $1/r^2$ suppression from the last term in brackets,
because the neutrino flux becomes
more collinear at large distances. The potential enters the Hamiltonian
as $v_r\,\mu(r){\bf D}$, where \be {\bf D}=\int_0^\infty
d\epsilon\left( {\bf P}(\epsilon)-\overline{{\bf
    P}}(\epsilon)\right)\,.  \ee Note that the Hamiltonian now depends
on ${\bf P}$ and $\overline{\bf P}$ themselves, thus making the flavor
evolution nonlinear.

The single-angle equations of motion for the flavor composition of neutrino and
antineutrino fluxes from a core-collapse supernova are then given by
\bea \frac{d{\bf
    P}(\epsilon)}{dr}&=&\left(+\frac{{\omega}(\epsilon){\bf
    B}}{v_r}+\frac{{\lambda(r)}{\bf L}}{v_r}+\mu(r){\bf
  D}\right)\times{\bf
  P}(\epsilon)\,,\quad\label{eom1}\\ \frac{d{\overline{\bf
      P}}(\epsilon)}{dr}&=&\left(-\frac{{\omega}(\epsilon){\bf
    B}}{v_r}+\frac{{\lambda(r)}{\bf L}}{v_r}+\mu(r){\bf
  D}\right)\times\overline{\bf P}(\epsilon)\,.\quad\label{eom2} 
  \eea
  
\subsection{Flavor Evolution}
\label{subsec:flavevol}

In the central regions of a core-collapse supernova, the matter
potential $\lambda(r)\gg\omega$, and the mixing angle $\theta_{13}$ is
suppressed by the factor
$\sim\omega/\lambda$~\cite{Wolfenstein:1977ue,Mikheev:1986gs}. As was
shown in ~\cite{Duan:2005cp, Hannestad:2006nj, Dasgupta:2007ws}, the
role of a large MSW potential is mimicked by setting the mixing angle
$\theta_{13}$ to a small value, and removing $\lambda(r)$ from the
equations of motion. We use this result without proof. Now, adding
Eq.\,({\ref{eom1}}) and Eq.\,({\ref{eom2}}) and integrating over all
energies, one finds that the vector $\int_0^\infty d\epsilon ({\bf
  P}+\overline{\bf P})-\frac{\widetilde{\omega}/v_r}{\mu}{\bf B}$
acts like a pendulum with the energy~\cite{Hannestad:2006nj,
  Duan:2007mv} \be E=\frac{\widetilde{\omega}}{v_r}{\bf
  B}\cdot\int_0^\infty d\epsilon\left({\bf P}+\overline{\bf
  P}\right)+\frac{1}{2}\mu(r)|{\bf D}|^2\,,\label{ener} \ee where
$\widetilde{\omega}$ is the the average of the oscillation frequency
$\omega$ over the spectrum of flux differences~\cite{Fogli:2007bk},
\bea \widetilde{\omega}&=&\frac{\int_{0}^{\infty} d\epsilon\,
  \omega(\epsilon)\left(\frac{d\Phi_{\nue}}{d\epsilon}-\frac{d\Phi_{\nux}}{d\epsilon}\right)}{2(\Phi_{\nue}-\Phi_{\nux})}\label{eq:ws}\\ &+&\frac{\int_{0}^{\infty}
  d\epsilon\,
  \omega(\epsilon)\left(\frac{d\Phi_{\nueb}}{d\epsilon}-\frac{d\Phi_{\nux}}{d\epsilon}\right)}{2(\Phi_{\nueb}-\Phi_{\nux})}\nonumber\,,
\eea and ${\bf B}$ is approximately equal to $\mp(0,\,0,\,1)+{\cal
  O}(\theta_{13}^2)$.

The dynamics of the neutrino flavor pendulum is approximately
determined by a comparison of the potential and kinetic energy of the
system~\cite{Hannestad:2006nj}. The potential energy of the system is
the first term on the r.h.s.\ of Eq.\,(\ref{ener}),
\ie~$\widetilde{\omega}{\cal F}_{+}/v_r$, where \be {\cal
  F}_{+}\equiv{\bf B}\cdot\int_0^\infty d\epsilon\left({\bf
  P}+\overline{\bf P}\right)\,, \ee which is initially \be {\cal
  F}_{+}(R_\nue)
=\mp\frac{\Phi_{\nue}+\Phi_{\nueb}-2\Phi_{\nux}}{\Phi_{\nue}+\Phi_{\nueb}+4\Phi_{\nux}}\,,\label{eq:Fplus}
\ee is the measure of the fraction of the neutrino flux available for
oscillation. We remind the reader that the $\mp$ sign depends on
whether $\dma>0$ (normal hierarchy) or $<0$ (inverted hierarchy). On
the other hand, the kinetic energy of the system is the second term on
the r.h.s.\ of Eq.\,(\ref{ener}), given by $\frac{1}{2}\mu(r){\cal
  F}_{-}^2$, where \be {\cal F}_{-}\equiv|{\bf D}|\,, \ee which is
initially \be {\cal F}_{-}(R_\nue)
=\frac{\Phi_{\nue}-\Phi_{\nueb}}{\Phi_{\nue}+\Phi_{\nueb}+4\Phi_{\nux}}\,,\label{eq:Fminus}
\ee is the net lepton asymmetry in the system.  For the first few
$100$~ms after bounce, in which the explosion mechanism must operate,
one typically has $\Phi_\nue > \Phi_\nueb > \Phi_\nux$, thus ${\bf
  P}_z$ and $\overline{\bf P}_z$ are positive except at the very
highest energies. For normal hierarchy, \ie~ $\dma>0$, the potential
energy $\widetilde{\omega}/v_r{\bf B}\cdot\int_0^\infty
d\epsilon\left({\bf P}+\overline{\bf P}\right)$ is already negative in
the initial state, and therefore the pendulum remains close to its
initial state (any other configuration would have higher energy). In
the inverted hierarchy, \ie~ $\dma<0$, however, the potential energy
is positive initially, and flipping the polarization vectors ${\bf P}$
and $\overline{\bf P}$ leads to a lowering of the total energy. This
instability of the neutrino flavor distribution leads to almost
complete flavor conversions by flipping the polarization vectors ${\bf
  P}$ and $\overline{\bf P}$ if $\dma<0$.

The above argument assumes that all ${\bf P}_z$ and $\overline{\bf
  P}_z$ are positive, which is true for supernova neutrino fluxes in
the early postbounce phase, at all but the very highest
energies. This may not always be the case and it is worth emphasizing
that even for normal hierarchy it is possible to get flavor flips, as first 
shown in~\cite{Dasgupta:2009mg}. However, that requires
a larger flux of $\mu$ and $\tau$ neutrinos (to make ${\bf P}_z$ and
$\overline{\bf P}_z$ negative) than is typically available in the
early postbounce phase.

When the kinetic energy $\frac{1}{2}\mu(r){\cal F}_{-}^2$ becomes
comparable to the potential energy $\widetilde{\omega}{\cal
  F}_{+}/v_r$, the flavor pendulum begins flipping back and forth
between the up and down states, and does not always return to its
initial position. This radius, below which the collective potential
pins all polarization vectors together and the motion is synchronized,
is given by the condition \be \mu(r_{\rm
  sync})\approx4\widetilde{\omega}\left(1+\frac{R_{\nue}^2}{4r_{\rm
    sync}^2}\right)\,\frac{{\cal F}_{+}}{{\cal
    F}^2_{-}}\,.\label{eq:muflip} \ee Beyond this radius, the neutrinos
begin to convert flavor as the flavor pendulum tends to drift towards
the lower energy configuration.

As the neutrinos stream out, the magnitudes of the vacuum Hamiltonian
$\widetilde{\omega}/v_r$ and the collective Hamiltonian $\mu|D|$
eventually become comparable. This happens at a radius $r_{\rm end}$
where $\mu(r_{\rm
  end})\approx\widetilde{\omega}\left(1+R_{\nue}^2/(4r_{\rm
  end}^2)\right)/{\cal F}_{-}$, the flavor pendulum settles into the
lower energy state, which involves a flip in flavor space in the
inverted hierarchy. Collective flavor conversions approximately
freeze-out at this radius. Vacuum and MSW neutrino oscillations
take place at much larger radii and we neglect them here.

The flip of the flavor pendulum, as described above, leads to a swap
of the $\nue$ and $\nueb$ number fluxes with those of $\nux$ and
$\nuxb$ number fluxes via pair-conversions
$\nue\nueb\leftrightarrow\nux\nuxb$. The number fluxes after
collective effects (on the l.h.s.) are given in terms of the number fluxes
before collective effects (on the r.h.s.) as \bea
&d\Phi_{\nue}/d\epsilon=\left\lbrace\begin{array}{c}d\Phi_{\nue}/d\epsilon\quad\quad\epsilon<\epsilon_{\rm
  split}\\ d\Phi_{\nux}/d\epsilon\quad\quad\epsilon>\epsilon_{\rm
  split}
\end{array}\right.&\,, \label{eq:swap1}\\
&d\Phi_{\nueb}/d\epsilon=d\Phi_{\nux}/d\epsilon&\,,\label{eq:swap2}\\ &4d\Phi_{\nux}/d\epsilon=d\Phi_{\nue}/d\epsilon+d\Phi_{\nueb}/d\epsilon+2d\Phi_{\nux}/d\epsilon&\label{eq:swap3}\,,
\eea where $\epsilon_{\rm split}$ is given by the
constraint~\cite{Raffelt:2007cb} \be \int_0^{\epsilon_{\rm split}}
d\epsilon\,d\Phi_{\nue}/d\epsilon=\int_0^\infty
d\epsilon\,\left(d\Phi_{\nue}/d\epsilon-d\Phi_{\nueb}/d\epsilon\right)\,.
\ee The sharp discontinuity for $\nu_e$ at $\epsilon_{\rm split}$ is
known in the literature as a spectral split, and appears at
\mbox{$\sim$$(6-8)$~MeV}. There is a spectral split in $\bar{\nu}_e$
too, but typically at even lower energies ($<5$~MeV) and
is generally ignored~\mbox{\cite{Fogli:2007bk, Dasgupta:2009mg}}.

\begin{figure}[!t]
\includegraphics[width=0.95\columnwidth]{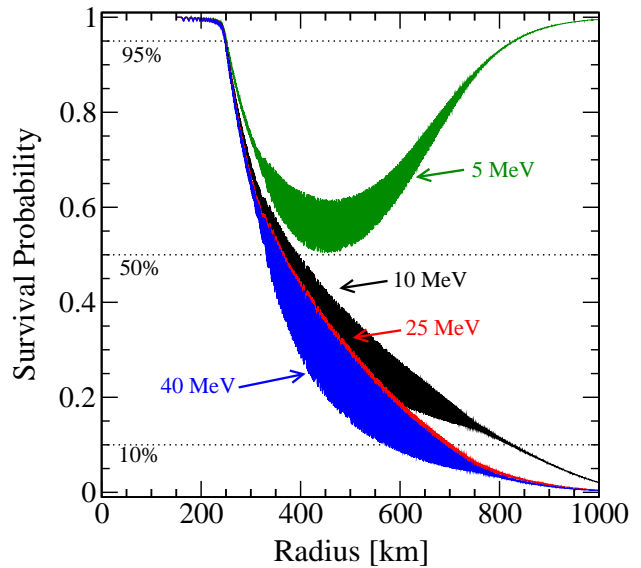}
\caption{Survival probabilities of flavor states $\nue$ (same for
  $\nux$) at representative energies (5, 10, 25, and 40 MeV) as a
  function of distance from the origin in a snapshot at 250 ms after
  bounce in the core collapse evolution of the $15$-$M_\odot$
  progenitor star discussed in Section~\ref{sec:models}.
\label{fig:flav-evol}}
\end{figure}

A representative example of $\nue$ flavor evolution is shown in
Fig.\,\ref{fig:flav-evol}. Note how the survival probability is
initially close to one ($\nue$ preserve their original state), begins
to decrease from $r_{\rm sync}$ (which in the example shown here is
$\sim250$~km), and finally around $r_{\rm end}$ (here $\sim700$~km)
asymptotes to zero (complete flavor conversion). Note that there are
neutrinos that return back to their original state -- those are the
lowest energy $\nue$ (below $\sim7$~MeV) that do not convert flavor as
mentioned above. We remind that in a two-flavor approximation, this
survival probability for $\nue$ is exactly the same as for $\nux$. The
behavior of $\nueb$ (and $\nuxb$) is only slightly different (energies
below $\sim4$~MeV return back to their original flavor), and therefore
not shown.

The value of the collective potential when the flavors flips start
occurring, \ie~$\mu(r_{\rm
  sync})=4\widetilde{\omega}\left(1+{R_{\nue}^2}/({4r_{\rm
    sync}^2})\right)\,{{\cal F}_{+}}/{{\cal F}^2_{-}}$,  does not
depend on the total neutrino number flux, but only on the relative
number fluxes. Note that $\Phi_{\nu,\nub}$ in $\mu(r)$ cancels 
with the denominator of ${\cal F}_-$, and is simply a choice of
normalization. With our normalizations for ${\bf P},\,\overline{\bf
  P}$, and $\mu(r)$, all model-dependence on neutrino spectra is
absorbed into one number, \ie~ $\widetilde{\omega}{\cal F}_+/{\cal
  F}_-^2$.
        
\begin{figure}[!t]
\vspace{0.07cm} 
\includegraphics[width=0.95\columnwidth]{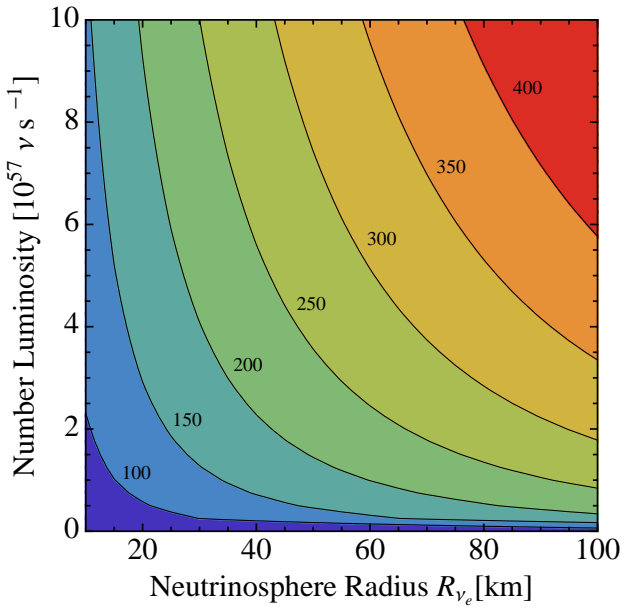}
\caption{Contours of equal synchronization radius (in km), using
  Eq.\,(\ref{eq:muflip}), as a function of the neutrino number
  luminosity ${\cal N}_{\nu,\nub}=4\pi R_{\nue}^2\Phi_{\nu,\nub}$ (in
  $10^{57}$~s$^{-1}$) and the $\nu_e$ neutrinosphere radius
  $R_{\nu_e}$ for a fiducial value of $\widetilde{\omega}{\cal
    F}_+/{\cal F}_-^2$ taken to be $50$~km$^{-1}$. See text for more
  details.}
\label{fig:rsyncontour}
\end{figure}

 For core-collapse supernova emission parameters predicted by typical
 simulations, $\widetilde{\omega}\,{{\cal F}_{+}}/{{\cal F}^2_{-}}$
 is typically in the range $(30-300)$~km$^{-1}$. We plot in
 Fig.\,\ref{fig:rsyncontour}, the synchronization radius $r_{\rm sync}$
 as a function of the number luminosity ${\cal N}_{\nu,\nub}$ and the
 neutrinosphere radius $R_{\nue}$ using Eq.\,(\ref{eq:muflip}) and assuming
 $\widetilde{\omega}\,{{\cal F}_{+}}/{{\cal
     F}^2_{-}}=50$~km$^{-1}$. Note that a factor of six increase in
 the value chosen as the critical $\mu(r_{\rm sync})$, reduces $r_{\rm
   sync}$ by only $\sim40\%$ because of the $1/r^4$ scaling of $\mu$.  
   In the postbounce pre-explosion phase, the
 total neutrino number luminosity is \mbox{${\cal
     N}_{\nu,\nub}\,$$\sim$$(1-10)\times10^{57}$~s$^{-1}$} and $\nu_e$
 neutrinosphere radius is~\mbox{$R_{\nue}\,$$\sim$$(40-80)$}~km, which leads to a typical
 synchronization radius \mbox{$r_{\rm sync}\,$$\sim$$(200-400)$~km}. Similarly $r_{\rm
   end}$ is seen to be $\sim$$(400-700)$~km.

Our discussion so far has been based on a single-angle formalism.
However, as we have already mentioned, neutrinos emitted at different
angles relative to the neutrinosphere experience different collective
and MSW potentials. This leads to multi-angle effects. There are
three ways in which these multi-angle effects are known to become
important.

First, if the MSW potential itself is much larger than the collective
potential weighted by the lepton asymmetry factor, \ie~
$\lambda(r)/v_r\gg\mu(r)|{\bf D}|$,  collective oscillations are
suppressed~\cite{EstebanPretel:2008ni}. Spelled out, this is the
case where
\begin{equation}
\lambda(r)\gg2\sqrt{2}G_F\Phi_{\nu,\nub}\frac{R_{\nue}^2}{r^2}\,{\cal F}_{-}\,.
\end{equation}

Second, if the $\nue$ and $\nueb$ fluxes are very similar,
\ie~ $\Phi_\nue\approx\Phi_{\nueb}$, a
single angle treatment is not appropriate due to multi-angle
decoherence~\cite{EstebanPretel:2007ec}. In this case, one finds that different
angular modes accrue random phases for both
normal and inverted hierarchy. Thus the polarization vectors
$\mathbf{P}$ and $\overline{\mathbf{P}}$ shrink to zero due to
kinematic decoherence. This can begin as soon as the synchronization
radius is reached and leads to rapid flavor equilibration (all flavors
assume the same spectrum and number flux).

Third, if the fluxes of $\mu$ and $\tau$ neutrinos and antineutrinos
become comparable to or greater than the $\nue$ or $\nueb$ fluxes, \ie~
$\Phi_{\nux}\gtrsim\Phi_{\nue},\Phi_{\nueb}$, there can be
self-induced suppression of collective
oscillations~\cite{Duan:2010bf}. If this were to happen, it
would delay flavor conversions to~larger radii. However, in the first
few $100$~ms after core bounce, $\Phi_{\nux}$ is generally
significantly smaller than $\Phi_{\nue}$ and $\Phi_{\nueb}$. Hence,
self-induced suppression of collective effects, shown to occur when
$\Phi_{\nux} \approx \Phi_{\nue} \approx \Phi_{\nueb}$
\cite{Duan:2010bf}, is rather unlikely~\cite{Fogli:2007bk}.
 
We do not have a detailed analytical understanding of these
multi-angle effects, and they will be studied numerically in the
Section~\ref{sec:maeffects}.  Before we perform a more detailed
numerical study, we can now provide first approximate answers to the
two questions raised at the end of Section~\ref{sec:revival}.

(i) \emph{What is the typical enhancement in heating that one can expect?} It
is easy to see that collective oscillations can give rise to almost
\emph{maximal} flavor conversion. All neutrinos and antineutrinos
change their spectra, thus the entire $\nue$ and $\nueb$ spectra can
get exchanged with those of $\nux$ leading to the largest possible
effect that can be expected from any flavor changing phenomenon. The
quantities responsible for heating, \ie~${\cal L}_{\nue}\langle
\epsilon^2_{\nue}\rangle$ and ${\cal L}_{\nueb}\langle
\epsilon^2_{\nueb}\rangle$, can get replaced by ${\cal
  L}_{\nux}\langle \epsilon^2_{\nux}\rangle$, which may be
significantly higher if the luminosities in all flavors are comparable
but $\nux$ energies are larger, leading to net enhancement of neutrino
heating.

(ii) \emph{Does this enhancement take place at sufficiently small
  radii to have a significant effect on the total neutrino heating}?
Based on typical postbounce neutrino emission characteristics, we
expect flavor exchange to begin at $r_{\rm
  sync}\,$$\sim$$(200-300)$~km and complete at $r_{\rm
  end}\,$$\sim$$(500-700)$~km at $\sim$$100\,\mathrm{ms}$ after
bounce. Neglecting multi-angle effects that could force $r_{\rm
  sync}\gtrsim700$~km \cite{Chakraborty:2011nf, Chakraborty:2011gd},
the oscillation radii will decrease in the later postbounce evolution,
since the neutrinospheres recede and the luminosities decrease with
time.  Hence, depending on the detailed dynamical evolution of a given
core collapse event, collective oscillations may indeed occur at
sufficiently small radii to significantly affect neutrino heating.  In
the next section, we will use full radiation-hydrodynamic
core-collapse supernova simulations to obtain a more quantitative
handle on the relevance of collective oscillations including
  multi-angle effects.
  
  \begin{figure*}[!t]
\includegraphics[width=0.87\textwidth]{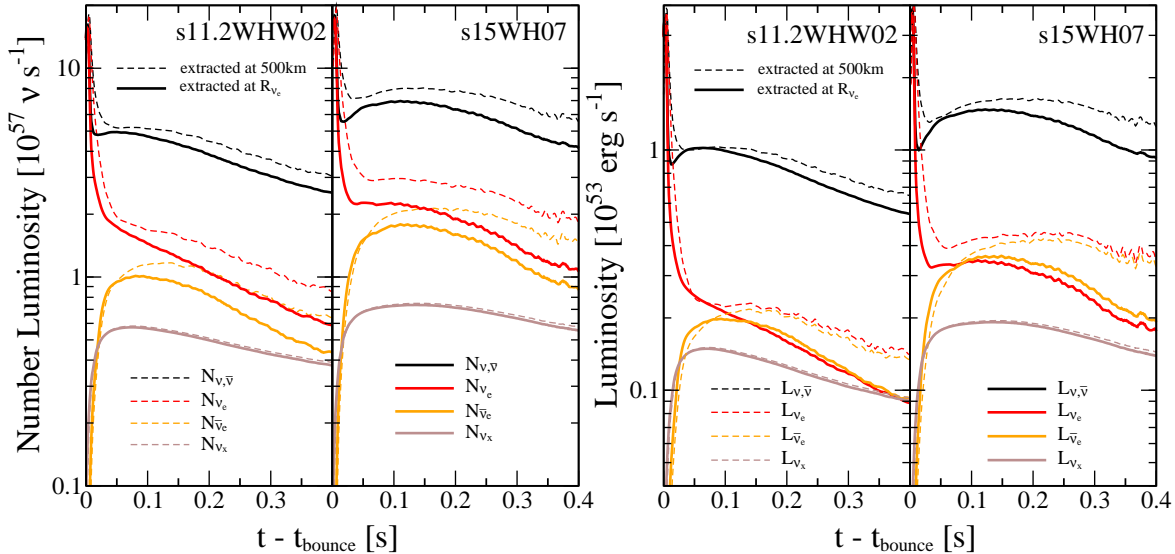}
\caption{Neutrino number luminosities (left plot) and energy
  luminosities (right plot) as a function of time after bounce for
  model s11.2WHW02 (left panels) and model s15WH07 (right panels).  We
  extract the neutrino luminosities at both the $\nue$ neutrinosphere
  and at 500~km.  The difference in the luminosities between $R_\nue$
  and 500~km is as expected: little difference in the $\nux$
  luminosities but significant differences in the $\nue$ and $\nueb$
  luminosities due to the large accretion luminosity, especially in
  model s15WH07.  We also show the total number luminosity, ${\cal
    N}_{\nu,\bar{\nu}} = {\cal N}_\nue + {\cal N}_\nueb + 4{\cal
    N}_\nux$, and the total energy luminosity ${\cal
    L}_{\nu,\bar{\nu}} = {\cal L}_\nue + {\cal L}_\nueb + 4{\cal
    L}_\nux$.}\label{fig:luminosities}
\end{figure*} 

\vspace{-0.3cm}
\section{Effect of Collective Oscillations on Supernova Shock Revival}
\label{sec:numerical}
To explore the potential effect of collective neutrino oscillations in
the core-collapse supernova environment more quantitatively, we
perform simulations with \code{VULCAN/2D}, an axisymmetric Newtonian
radiation-hydrodynamics code
\cite{livne:93,livne:07,burrows:07a,ott:08}.  In the variant of
\code{VULCAN/2D} that we use here, neutrino transport is handled in
the multi-group flux-limited diffusion (MGFLD) approximation to the
full Boltzmann equation, evolving the mean radiation intensity and
using Bruenn's flux limiter~\cite{bruenn:85}.  Velocity dependence and
energy-redistribution between neutrino groups via inelastic scattering
are neglected.  Three neutrino species are evolved, $\nue$ and $\nueb$
along with a representative $\mu,\tau$ flavor neutrino, $\nux$, using
16 energy groups, logarithmically spaced from $2.5$ to
$250\,\mathrm{MeV}$.  The neutrino opacities are taken from
\cite{brt:06}.  More details on \code{VULCAN/2D} are provided in
\cite{burrows:07a,ott:08,livne:07}.

The computational grid consists of an inner quasi-Cartesian region
that gradually transitions to an outer polar grid starting at a radius
of $20\,$km and extending out to $5000\,\mathrm{km}$ with $221$
logarithmically-spaced radial zones and $121$ angular zones, covering
the full $180^\circ$ of the axisymmetric domain.  Using a Cartesian
center avoids the small time steps associated with converging angular
zones of a polar grid. See Fig.\,4 of \cite{ott:04} for an example of
our grid setup.

It is obvious from Fig.\,\ref{fig:rsyncontour} that low neutrino number
luminosities are favorable for collective oscillations to occur at
small radii where they may have an impact on neutrino heating.  There
is a general trend (at least in the $\sim(10-20)$-$M_\odot$ ZAMS mass
range) for more massive progenitors to lead to higher postbounce
neutrino luminosities \cite{buras:06a}. Hence, in this study, we
perform calculations with the nonrotating $11.2$-$M_\odot$ progenitor
model of Woosley~\etal~\cite{whw:02} (solar composition, model
s11.2WHW02 in the following) and, for comparison, also with the
nonrotating $15$-$M_\odot$ progenitor model of Woosley \&
Heger~\cite{woosley:07} (also solar composition, model s15WH07 in the
following).

We evolve both progenitors with the H. Shen~\emph{et al.} EOS (HShen
EOS, \cite{shen:98a,hshen:11}), which is based on a relativistic mean
field model of nuclear matter and yields a maximum gravitational mass
of $2.24$-$M_\odot$ for a cold neutron star.  We follow models
s11.2WHW02 and s15WH07 from the onset of core collapse to
$400\,\mathrm{ms}$ after core bounce and do not observe an onset of
explosion in either model before we terminate our calculations.  For
the $11.2$-$M_\odot$ progenitor, Buras {\emph{et al.}}~\cite{buras:06b} observed an early and weak neutrino-driven,
SASI-aided explosion. For the $15$-$M_\odot$ progenitor,
Bruenn~\emph{et al.}~reported an explosion setting in at $\sim
300\,\mathrm{ms}$ after bounce.  These differences in outcome observed
by these groups may be due to their use of the softest variant of the
Lattimer-Swesty EOS \cite{lseos:91} (which has now been ruled out
\cite{demorest:10}), inclusion of general-relativistic effects and/or
their more sophisticated treatment of neutrino transport.

\vspace{-0.4cm}
\subsection{Postbounce Evolution:\\ Neutrino Radiation Fields and
Hydrodynamics}
\label{sec:models}

In Fig.\,\ref{fig:luminosities}, we show the postbounce neutrino energy
luminosities (${\cal L}_{\nu_i}$) and number luminosities (${\cal
  N}_{\nu_i}$) for models s11.2WHW02 and s15WH07.  We extract the
angle-averaged luminosities both at the $\nue$ neutrinosphere
($R_\nue$) and at a radius of 500$\,$km.  Significant $\nue$ and, to
some extent, $\nueb$ emission does occur outside the neutrinosphere
due to charged-current interactions involving accreted dissociated
material (\emph{accretion luminosity}; \emph{e.g.},
\cite{fischer:09a}). Since the $\nux$ do not participate in
charged-current interactions, their luminosities evolve little between
$R_\nue$ and 500$\,$km.

In model s15WH07, the neutrino luminosities are consistently
higher than in model s11.2WHW02.  This is due to the higher
temperatures in this progenitor, which lead to higher core
luminosities, and to a higher accretion rate, which leads to higher
accretion luminosities.  With the exception of a very short period
close to bounce ($<30\,$ms), the standard hierarchy of neutrino number
flux ($\Phi_\nui = {\cal N}_\nui / 4\pi r^2$), $\Phi_\nue > \Phi_\nueb
> \Phi_\nux$, is achieved in both models at $R_\nue$ and at 500~km.
Such a hierarchy is not present in the energy luminosities at
$R_\nue$, but is obtained when taking the accretion luminosity into~account. 

In the following, we make the simplifying assumption that neutrinos
traveling through the neutrinosphere will undergo collective
oscillations but those emitted as part of the accretion luminosity
will not.  This approximation is difficult to overcome, since a full
collisional Boltzmann solution including collective oscillations would
be required for a self-consistent treatment. However, tests in which
we used the asymptotic (${\cal L}_{\nu_i}^\mathrm{ns}$ + ${\cal
  L}_{\nu_i}^\mathrm{acc}$) instead of the neutrinospheric
luminosities led to no qualitative and only small quantitative
differences in the critical oscillation radii.

\begin{figure*}[!htbp]
\includegraphics[width=0.45\textwidth]{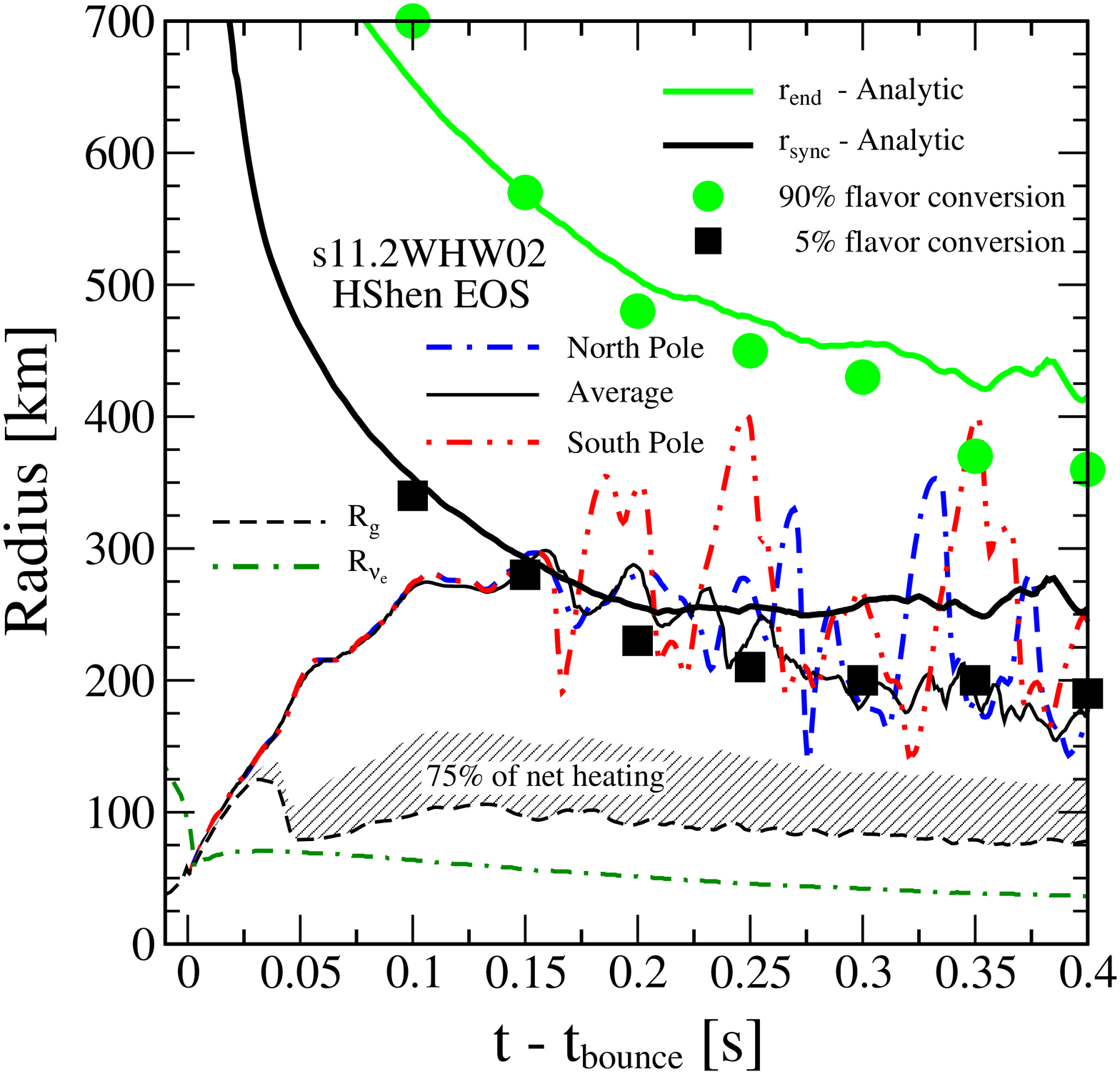}
\includegraphics[width=0.437\textwidth]{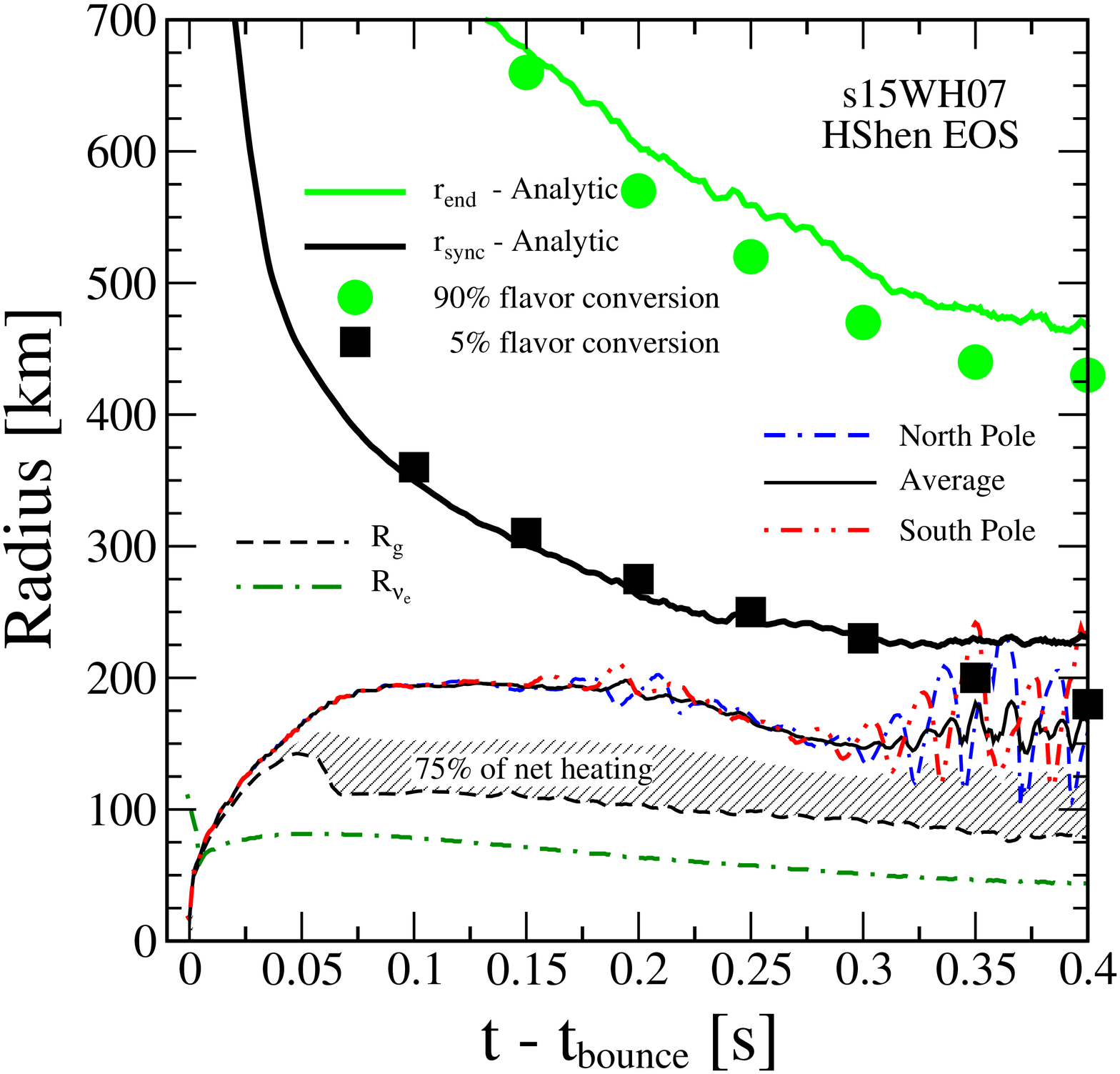}
\caption{Comparison of neutrino oscillation radii to shock 
    and gain radii.  Average (thin solid lines) and polar shock radii
  (dot-dot-dashed and dot-dashed-dashed lines) in the postbounce phase
  of model s11.2WHW02 (left plot) and model s15WH07 (right plot). Also
  shown (dot-dashed lines) are the energy-averaged $\nu_e$
  neutrinosphere locations and the angle-averaged gain radii (dashed
  lines). The hashed region just above the gain radius denotes the
  area in which 75\% of the net heating occurs. We show the radii at
  which collective neutrino oscillations begin ($r_\mathrm{sync}$;
  dark black lines and symbols) and end ($r_\mathrm{end}$; light green
  lines and symbols) as obtained via an analytic approximation (solid
  lines) and via detailed numerical calculations at select times
  (squares and circles).  $r_\mathrm{sync}$ is initially large, but
  drops to radii comparable to the shock radii at $\sim$$(100-150)\,$ms
  in model s11.2WHW02 and at $\sim$$(300-350)\,$ms in model
  s15WH07. See text for details.}
\label{fig:s11.2_and_s15}
\end{figure*}

In the left panel of Fig.\,\ref{fig:s11.2_and_s15}, we show the
evolution of the shock radii at the North and South pole, and of the
angle-averaged shock radius in model s11.2WHW02. Also shown are the
energy-averaged $\nue$ neutrinosphere and gain radii in this
model. Until $\sim$$150\,$ms after bounce, the shock remains
essentially spherically symmetric. Then, the SASI begins to grow and
large asymmetries arise in the shock front.  The $\ell=1$ sloshing of
the shock radius in the North-South direction is characteristic of the
SASI in 2D (cf.\ \cite{blondin:03}, but note that 3D gives a different
SASI behavior, \emph{e.g.},~\cite{nordhaus:10}).  While oscillations
in the shock position reach radii upwards of $\sim$$400\,$km, the
average peaks at $\sim$$300\,$km at $\sim$$150\,$ms after bounce, then
slowly recedes, reaching $\sim$$200\,$km at $350\,$ms.  The
angle-averaged gain radius, where charged-current neutrino heating
balances cooling, hovers around $\sim$$100\,$km.  The hashed region
denotes the angle-averaged radial extent in which 75\% of the net
heating occurs. The charged-current interactions are most effective at
transferring energy to the matter at high density
(cf.~Eq.\,(\ref{eq:heating_one})), therefore, most of the net heating
occurs near the gain radius where the density is the highest 
(cf.~the discussion in \cite{janka:01} and Fig.~10 of \cite{ott:08} depicting
the heating rate as a function of radius). The energy-averaged $\nue$
neutrinosphere radius peaks near $\sim$$70\,$km at $\sim$$40\,$ms
after bounce and recedes thereafter. By $350\,$ms after bounce, the
$\nu_e$ neutrinosphere has receded to $\sim$$40\,$km.

The postbounce evolution of model s15WH07 is summarized by the right
panel of Fig.\,\ref{fig:s11.2_and_s15}.  The postbounce dynamics is
qualitatively similar to model s11.2WHW02's and quantitative
differences are due primarily to model s15WH07's higher postbounce
accretion rate. This prevents the shock from reaching the higher radii
achieved in model s11.2WHW02 before stagnation and suppresses the
development and the strength of the SASI until later~times.

\vspace{-0.4cm}
\subsection{Collective Neutrino Oscillation Radii}
\label{sec:oscirad}

Collective neutrino oscillations may be relevant in the postbounce
evolution if they occur within the region behind the shock where
conditions are conducive to net neutrino heating. We
post-process the spectral neutrino fluxes predicted by the {\tt
  VULCAN/2D} simulations in two ways to determine the radius at which
collective oscillations may begin.  As input to these calculations, we
chose the neutrino spectra at the $\nu_e$ neutrinosphere, which we
average over lateral angle\footnote{Since our models are nonrotating, the
variation of the neutrino spectra with lateral angle is not large
(cf.~\cite{ott:08,brandt:11,marek:09b,lund:10}).}.

Our first method is using the analytic expressions of Section
\ref{collective}.  We invert Eq.\,(\ref{eq:muflip}) using
Eqs.\,(\ref{eq:muofr}), (\ref{eq:ws}), (\ref{eq:Fplus}), and
(\ref{eq:Fminus}) and solve for $r_\mathrm{sync}$,
\begin{equation}
 r_\mathrm{sync}=R_\nue \left[
   \frac{1+\sqrt{2}G_F\Phi_{\nu,\bar{\nu}}/(\widetilde{\omega}\mathcal{F}_+/
     \mathcal{F}^2_{-})}{\sqrt{9+8\sqrt{2} G_F\Phi_{\nu,\bar{\nu}}
       /(\widetilde{\omega} \mathcal{F}_+/ \mathcal{F}^2_{-})}-1}
   \right]^{1/2}\,.\label{eq:rsync}
\end{equation}
We also use $\mu(r_{\rm
  end})\approx\widetilde{\omega}\left(1+R_{\nue}^2/(4r_{\rm
    end}^2)\right)/{\cal F}_{-}$ to obtain an estimate for the radius
at which the oscillations effectively are complete.  Once again
inverting this using Eqs.\,(\ref{eq:muofr}), (\ref{eq:Fminus}), and (\ref{eq:ws}), we obtain
for $r_\mathrm{end}$,
\begin{equation}
 r_\mathrm{end}=R_\nue \left[
   \frac{1+4\sqrt{2}G_F\Phi_{\nu,\bar{\nu}}\mathcal{F}_{-}/\widetilde{\omega}}{\sqrt{9+32\sqrt{2} G_F\Phi_{\nu,\bar{\nu}} \mathcal{F}_{-}
       /\widetilde{\omega}-1}}
   \right]^{1/2}\,.\label{eq:rend}
\end{equation}

As an alternative to the rather rough approximation of
Eqs.\,(\ref{eq:rsync}) and (\ref{eq:rend}), we determine the critical
collective neutrino oscillation radii by numerically solving the set
of coupled nonlinear differential Eqs.\,(\ref{eom1}) and (\ref{eom2})
with the initial conditions given by Eqs.\,(\ref{Pvec}) and
(\ref{Pvecb}) based on the neutrino number fluxes at $R_{\nu_e}$ in
the {\tt VULCAN/2D} simulations. The equations are solved as a function of 
radius $r$ for 32, 64, and 128 energy groups spaced as in Gauss-Legendre quadrature with the
oscillation code of Dasgupta~\emph{et al.}~\cite{Dasgupta:2007ws} in
combination with the \mbox{open-source} ordinary differential equation solver
{\tt CVODE}~\cite{CVODE}.  We carry out
  this calculation for select postbounce times and numerically
  identify $r_{\rm sync}$ and $r_{\rm end}$ with the radii at which
  $5\%$ and $90\%$ flavor conversion have occurred, respectively.  For
  reference, Fig.\,\ref{fig:flav-evol} shows the $\nue$ survival
  probability for select energies obtained with such an evolution for
  the neutrino spectra in model s15WH07 at 250~ms after bounce.

We present the results of both methods applied to models 
s11.2WHW02 and s15WH07 in Fig.\,\ref{fig:s11.2_and_s15}.
The values of $r_\mathrm{sync}$ and $r_\mathrm{end}$ predicted
 by the two methods agree well at early to intermediate times.
At late times, the analytic approximation overpredicts by 
$\sim$$25\%$. Note that, although the analytical formulae derived 
in Sec.\,\ref{subsec:flavevol} are in a two-flavor approximation, 
they can be compared to a three-flavor numerical calculation, 
because the third flavor is almost decoupled.

It is apparent from Fig.\,\ref{fig:s11.2_and_s15} that the radial
interval over which the collective oscillations occur is well outside
the shock at early times. Only after
$\sim$$150\,\mathrm{ms}$ ($\sim$$350\,\mathrm{ms}$) in model s11.2WHW02
(s15WH07), when the total number luminosity has decreased and the
neutrinosphere has receded, does $r_\mathrm{sync}$ recede below the
shock radius and collective oscillations can have an effect on the
subsequent evolution.

\vspace{-0.4cm}
\subsection{Multi-angle Effects}
\label{sec:maeffects}
To ascertain the importance of multi-angle effects we use the neutrino
luminosities and electron density profile as predicted along the
equatorial direction by our {\tt VULCAN/2D} simulations. Due to the
high computational demand of multi-angle oscillation calculations, we
do not include the energy spectra of the different flavors, and
instead assume a monoenergetic ensemble with the average vacuum
oscillation frequency $\widetilde{\omega}$. This is the same
approximation as used by \cite{Chakraborty:2011nf,Chakraborty:2011gd}.
Since our \code{VULCAN/2D} simulations made use of the efficient MGFLD
variant of the code, which does not carry direct information on the
momentum-space angle dependence of the neutrino radiation field, we
compare with the fully angle-dependent calculations of \cite{ott:08}
and construct approximate angle-dependent radiation fields. We find
that the angular distribution of the neutrino luminosity, derived from
the simulations of \cite{ott:08}, is parametrized quite well by
$d\Phi/d\cos\vartheta_{R_{\nue}}\propto\exp\left[(\cos\vartheta_{R_{\nue}}-1)/\sigma_{R_{\nue}}\right]$,
where we choose $\sigma_{R_\nue} = 0.357(\langle 1/\mathcal{F}\rangle
-1)$, and where $\langle 1/\mathcal{F}\rangle$ is the inverse flux
factor. This parameterization reproduces both the isotropic ($\langle
1/\mathcal{F}\rangle \gg 1$, $\sigma \gg 1$, and $d\Phi/d\cos\vartheta
\propto$ constant) and the free streaming ($\langle
1/\mathcal{F}\rangle \sim 1$, $\sigma \sim 0$, and
$d\Phi/d\cos\vartheta \propto \delta(\vartheta)$) limits, and
qualitatively reproduces the angular distribution at the
neutrinosphere (cf. Figure 3 of \cite{ott:08}).  At the
neutrinosphere, our inverse flux factors are $\sim$4-5. We assume a
sharp neutrinosphere, cutting off all neutrinos traveling backwards
into the neutrinosphere. With these choices, we solve
Eqs.~(\ref{eom-ma1}) and (\ref{eom-ma2}) using a multi-angle
oscillation code that is technically similar to the one used for the
single-angle calculations.{\footnote{We have verified numerical convergence by comparing results from calculations with 100, 200, and 400 angular bins. The local error tolerance was fixed at $10^{-12}$, which allows us to achieve convergence with $\sim400$ modes. Additionally, we have verified the calculations with 800 modes in a limited number of cases, and found them to be consistent. We have also reproduced results similar to Ref.\cite{EstebanPretel:2007ec}.}.

\begin{figure*}[!thpb]
\includegraphics[width=0.93\textwidth]{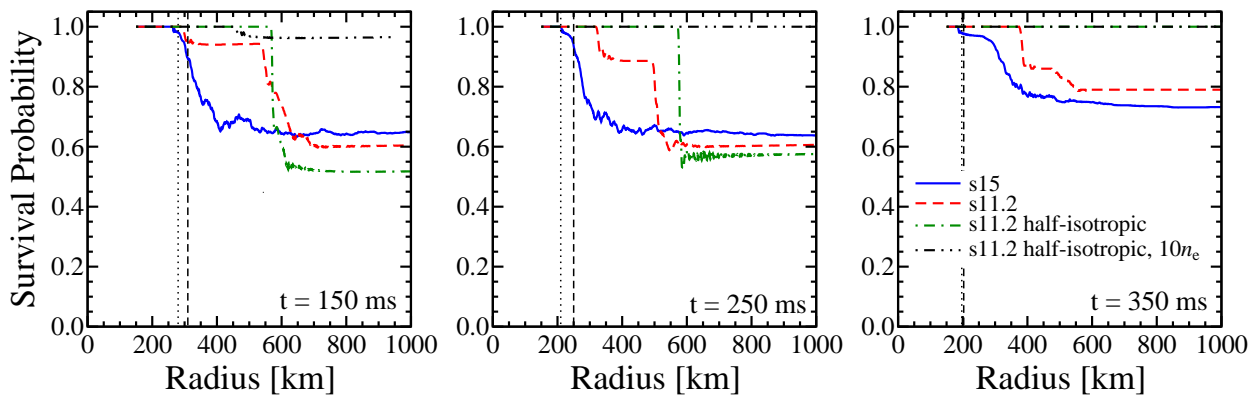}
\caption{Survival probabilities (averaged over all emission angles)
  calculated using a multi-angle oscillation code for a single average
  energy. The different panels correspond to different times for both
  models -- $t=150\,$ms~(left), $t=250\,$ms~(center), and
  $t=350\,$ms~(right). Blue solid lines denote results for model
  s15WH07 and dashed red lines are results for model s11.2WHW02.  We
  also plot results of our cross-checks: dash-dotted green lines are
  results for model s11.2WHW02 with half-isotropic angular emission
  and dot-dot-dashed black lines show for the same model the case with
  half-isotropic angular emission and 10 times higher electron
  density.  The vertical lines are from our {\emph{single angle}}
calculations and represent the radii at which 5\% flavor conversion
has occurred in models s11.2WHW02 (dotted) and s15WH07
  (dashed).}
\label{fig:s11.2_and_s15-MA-2}
\end{figure*}

In Fig.~\ref{fig:s11.2_and_s15-MA-2}, we show the survival probability
(averaged over all emission angles) of the average-energy neutrinos
calculated using the multi-angle code. The three panels show the
situation at $150\,$ms~(left), $250\,$ms~(center), and
$350\,$ms~(right) after core bounce and results for model s11.2WHW02
and s15WH07 are shown in dashed red and solid blue lines,
respectively.  The vertical lines in Fig.~\ref{fig:s11.2_and_s15-MA-2}
are from our {\emph{single angle}} calculations and represent the
radii at which 5\% flavor conversion has occurred for the s11.2WHW02
(dotted) and the s15WH07 (dashed) progenitors.

We find that in model s11.2WHW02 the onset of flavor conversion is
delayed by multi-angle effects to $\sim~(300-500)$\,km. This is
expected and due to high electron number density.  As we remarked in
Section\,{\ref{collective}}, whenever the MSW potential $\lambda(r)\gg
\lambda_{\rm MA}(r)=2\sqrt{2}G_F\Phi_{\nu,\nub}(R^2_{\nue}/r^2) {\cal
  F}_{-}$, multi-angle effects are strong and suppress the
oscillations. In Fig.~\ref{fig:s11.2_and_s15-MA}, we plot $\lambda(r)$
and $\lambda_\mathrm{MA}(r)$ as a function of radius at 150\,ms (left
panels), 250\,ms (center panels), and 350\,ms (right panels) after
bounce, for models s11.2WHW02 (top panels) and s15WH07 (bottom
panels). Radial profiles along ten lateral directions from the North
to South pole are shown to capture variations in $\lambda(r)$ due to
SASI oscillations.  $\lambda(r)$ is almost always larger than
$\lambda_{\rm MA}$, which decreases $\propto r^{-2}$ as
expected. $\lambda(r)$, which effectively traces the electron number
density, falls off $\propto r^{-3}$ below the shock radius.  Above the
shock, where matter is essentially in free fall, the density roughly
follows $r^{-1.5}$. Typically, the ratio $\lambda(r)/\lambda_{\rm
  MA}(r)$ is in the range $(1-100)$, getting close to 1 at later times
($t\gtrsim 250$\,ms) at $r\sim$$200$\,km in model~s11.2WHW02.
  
  \begin{figure*}[!thpb]
\includegraphics[width=0.93\textwidth]{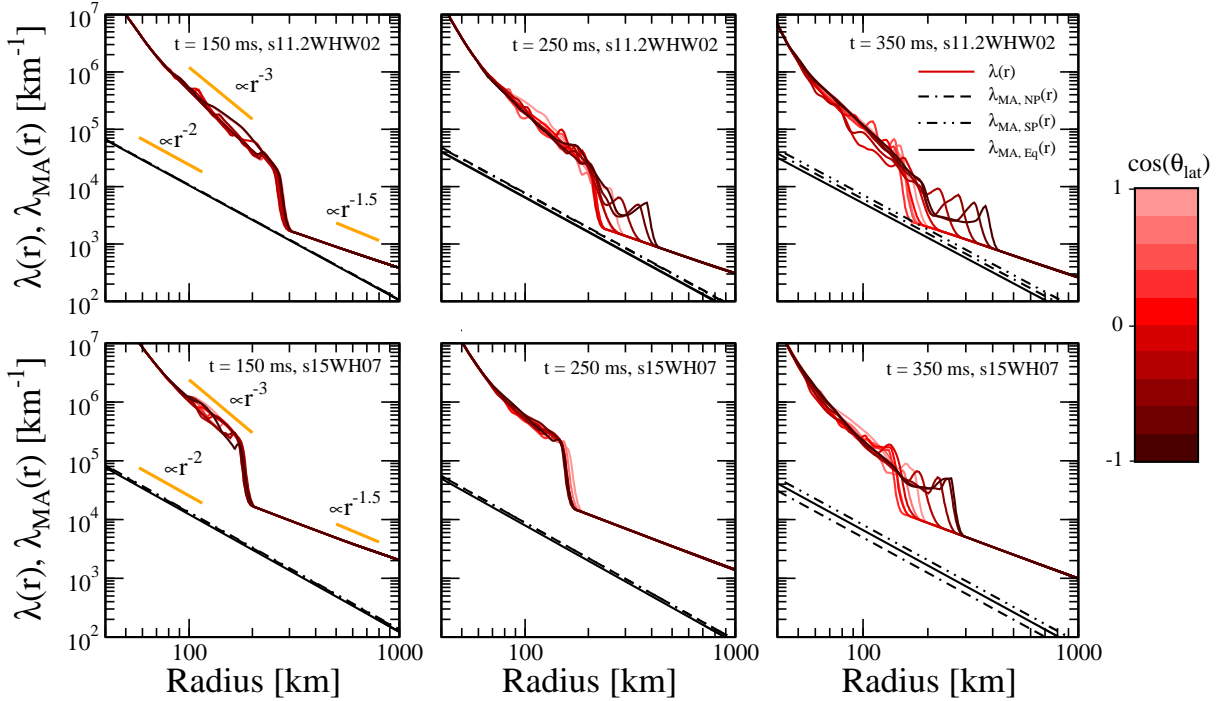}
\caption{The MSW potential $\lambda(r)$ along various directions (thin
  solid lines, 10 rays equally spaced in
  $\cos({\theta_\mathrm{lat})}$), in comparison to the minimum
  $\lambda(r)$ needed for multi-angle suppression $\lambda_{\rm
    MA}=2\sqrt{2}G_F\Phi_{\nu,\nub}(R^2_{\nue}/r^2 ){\cal F}_{-}$
  (thick dot-dashed line, taken along the North pole (NP), South pole
  (SP), and equator (EQ)). To guide the eye, rulers show $r^{-1.5},\,
  r^{-2},$ and $r^{-3}$ radial profiles. The steep rise in the
  $\lambda (r)$ profiles occurring around $r \sim$$200\,$km, but
  varying with lateral angle, is the location of the shock. The MSW
  potential $\lambda(r)$ is generally larger than the critical value
  $\lambda_{\rm MA}(r)$ needed for multi-angle suppression of neutrino
  oscillations.}
\label{fig:s11.2_and_s15-MA}
\end{figure*}

 Comparing $\lambda(r)$ along different directions, we find that the
 SASI oscillations developing at $t-t_{\rm bounce} \gtrsim150$\,ms
 lead to a significant spread in the values of $\lambda(r)$,
 sometimes, in model s11.2WHW02, bringing it just below the critical
 value below which multi-angle suppression is less effective. This
 occurs when the shock recedes to particularly small radii and the
 region in which the suppression is lifted is always at or outside the
 shock. 
   
On the other hand, as we remarked in Section~\ref{subsec:flavevol},
multi-angle effects may appear in a second way: If the $\nue$ and
$\nueb$ fluxes are very similar, \ie~ $\Phi_\nue\approx\Phi_{\nueb}$,
multi-angle decoherence of flavors sets in. In our model
s11.2WHW02, the $\nue$ and $\nueb$ fluxes are not too similar
($(\Phi_\nue-\Phi_\nueb)/\Phi_\nueb\approx0.25$), thus the decoherence
effect remains negligible.  
   
In model s15WH07, on the other hand, the multi-angle oscillation
calculation does not predict a significant delay in radius compared to
the single-angle prediction, in spite of the high electron number
density also present in this model.  Since in model s15WH07 the $\nue$
and $\nueb$ fluxes are quite similar
($(\Phi_\nue-\Phi_\nueb)/\Phi_\nueb\approx0.1$) early on, we attribute
this surprising result to (at least partial) multi-angle flavor
decoherence, which leads to oscillations despite the high electron
number density. We find flavor conversion almost as soon as the
neutrinos cross the synchronization radius. This observation is
consistent with the previous work by Esteban-Pretel~\etal, in which
where they found that $(\Phi_\nue-\Phi_\nueb)/\Phi_\nueb\lsim0.2$
leads to decoherence~\cite{EstebanPretel:2007ec}.

Comparing our results quantitatively with the recent multi-angle work
of Chakraborty~et~al.~\cite{Chakraborty:2011nf, Chakraborty:2011gd} is
not straightforward, since these authors based their calculations on
1D supernova simulations using a different progenitor model (a
$10.8$-$M_\odot$ progenitor of \cite{whw:02}).  However, we note that
they observe a stronger multi-angle suppression than borne out by our
models. For example, flavor conversion is delayed for their
$10.8$-$M_\odot$ progenitor almost until $(700-1000)$\,km, while we
observe flavor evolution to begin around already at $(300-500)$\,km in
our models. We believe that this is due to two reasons:
 
 Firstly, a half-isotropic angular distribution
 ($d\Phi/d\cos\vartheta_{R_{\nue}}\propto\cos\vartheta_{R_{\nue}}$)
 was used in the calculations of~\cite{Chakraborty:2011nf,
   Chakraborty:2011gd}. Compared to our angular distribution, this
 underestimates neutrinos emitted at large angles. The collective
 interaction is stronger for tangentially emitted neutrinos than for
 radially emitted neutrinos, therefore suppressing the tangential
 modes leads to a slower growth of the collective instability. We
 verify this claim by replacing the angular distribution in our model
 s11.2WHW02 by a half-isotropic angular distribution. The results for
 this half-isotropic case are shown in
 Fig.~\ref{fig:s11.2_and_s15-MA-2} (dot-dashed green line). The ``double-step'' feature only appears 
when we use our angular emission spectrum, which introduces stronger collective effects. This suggests that the feature is related to the angular spectrum, and not a numerical artifact. It is thus
 clear that simply changing the angular distribution can change the
 onset of flavor conversion. This may be strong enough to create a
 qualitative difference as evident from the snapshot at $t=350$\,ms in
 our $11.2$-$M_\odot$ model, where we find that the oscillations do
 not occur for a half-isotropic angular distribution, but do occur for
 our angular distribution modeled after the full 2D multi-angle
 neutrino transport simulations of \cite{ott:08}.
 
Secondly, in the $10.8$-$M_\odot$ model of ~\cite{Chakraborty:2011nf,
  Chakraborty:2011gd}, one finds a ratio of electron to neutrino
density that is up to $10$ times larger than in our $11.2$-$M_{\odot}$
model at various radii and times. This is due to the different
progenitor structure used -- our s11.2WHW02 model has a lower
postbounce accretion rate than the models of \cite{Chakraborty:2011nf,
  Chakraborty:2011gd}. We verify that this is indeed an important
factor, by artificially increasing the electron density by a factor of
10 and replacing the angular distribution in our oscillation
calculations by a half-isotropic angular spectrum for model
s11.2WHW02. This case is expected to closely follow the results
of~\cite{Chakraborty:2011nf, Chakraborty:2011gd}.  The results of this
are shown in Fig.~\ref{fig:s11.2_and_s15-MA-2} (dot-dot-dashed black
line). They demonstrate that such a change in the electron density and
angular distribution can indeed significantly suppress the flavor
evolution, in agreement with previous
results~\cite{Chakraborty:2011nf, Chakraborty:2011gd}.

In the light of these results, we believe that the role of multi-angle
effects remains an issue that requires further scrutiny. The role of
the matter density, $\nue/\nueb$ asymmetry, and the angular
distribution need to be studied in more detail. Flavor conversion are
not always completely suppressed due to multi-angle matter
suppression. Predictions of the neutrino flavor content at early times
must therefore be used with abundant caution. Fortunately, as we shall
see in the next section, our conclusions regarding the impact of
collective oscillations on the supernova mechanism remain largely
unchanged.
  
\subsection{Potential Enhancement of Neutrino Heating}
  
As discussed in Section~\ref{collective}, collective neutrino
oscillations (in the inverted mass hierarchy) will lead to a swap of
the $\nue$ and $\nueb$ spectral fluxes with the spectral fluxes of the
$\nu_x$ neutrinos (cf.~Eqs.\,(\ref{eq:swap1})-(\ref{eq:swap3})).  In
order to illustrate the potential enhancement of neutrino heating due
to this swap, we first consider the original heating rate before
oscillations (cf.~Eq.\,(\ref{eq:heatrate})), which can be expressed as
\begin{eqnarray}
  \nonumber \mathcal{H}_\mathrm{before} &\sim& \hat{\sigma} \langle
  \epsilon^2_{\nue} \rangle [{\cal L}_{\nue}^\mathrm{ns} + {\cal
      L}_{\nue}^\mathrm{acc}]c_N \\ 
  &&\hspace*{1.0cm} +\ \hat{\sigma}
  \langle \epsilon^2_{\nueb} \rangle [{\cal L}_{\nueb}^\mathrm{ns} +
    {\cal L}_{\nueb}^\mathrm{acc}]c_P\,,\label{eq:heat_before}
\end{eqnarray}
where we explicitly split the luminosity into
  core luminosity emanating from the neutrinosphere
  (${\cal L}_\nui^\mathrm{ns}$) and accretion luminosity
(${\cal L}_\nui^\mathrm{acc}$), the latter being emitted
  almost entirely interior to the gain region. We take $\langle
  \epsilon^2_{\nu_i} \rangle$ as the value calculated at the $\nu_e$
  neutrinosphere. The $\nu_x$ do not take part in charge-current
interactions and play no role in the heating before oscillations.

We now estimate the heating rate after taking into account a partial
conversion of the neutrino spectra in the region behind the shock, as
observed at late times in our simulations. 
\begin{eqnarray}
\nonumber
 \mathcal{H}_\mathrm{after} & \sim & \mathcal{H}_\mathrm{before} \\
\nonumber
&& \hspace*{0.5cm} +\ \hat{\sigma}\left[\langle \epsilon^2_{\nux} \rangle \mathcal{L}^\mathrm{ns}_\nux - \langle \epsilon^2_{\nue} \rangle \mathcal{L}^\mathrm{ns}_\nue\right] c_N^O \\
\nonumber
&& \hspace*{0.5cm} +\ \hat{\sigma}\left[\langle \epsilon^2_{\nux} \rangle \mathcal{L}^\mathrm{ns}_\nux - \langle \epsilon^2_{\nueb} \rangle \mathcal{L}^\mathrm{ns}_\nueb\right] c_P^O \\
&& \hspace*{0.5cm} +\ \hat{\sigma}\langle \epsilon^2_{\nue} \rangle^\star \left[\langle \epsilon_\nue \rangle^\star ({\cal N}^\mathrm{ns}_\nue - {\cal N}^\mathrm{ns}_\nueb \right)] c_N^O\,.\label{eq:heat_after}
\end{eqnarray}
The stars $(\star)$ on $\langle \epsilon_{\nue}^2\rangle$ and $\langle
\epsilon_{\nue}\rangle $ denote that the respective averages are taken
over only the part of the spectrum below the split energy
$\epsilon_{\rm split}$ (cf.~Section~\ref{collective}).  $c_N^O$ and
$c_P^O$ are the column number densities of interacting baryons taking
oscillations into account,
\begin{equation}
c_i^O = \int_{R_\mathrm{g}}^{R_\mathrm{s}} dr\ n_i\ P_\mathrm{ex}(r)\,,
\end{equation}
where $P_\mathrm{ex}(r)$ is the flavor conversion fraction and
$R_\mathrm{g}$ and $R_\mathrm{s}$ are the gain and shock radius,
respectively.  If no oscillations occur, $P_\mathrm{ex} = 0$
everywhere and $c_i^O$ is zero, $\mathcal{H}_\mathrm{after} =
\mathcal{H}_\mathrm{before}$.  If complete oscillations occur before
$R_\mathrm{g}$, $P_\mathrm{ex} = 1$ everywhere and $c_i^O = c_i$. All
other quantities in Eq.\,(\ref{eq:heat_after}) have their
pre-oscillated values. 

\begin{figure}[t]
\vspace{-0.3cm}
\includegraphics[width=0.95\columnwidth]{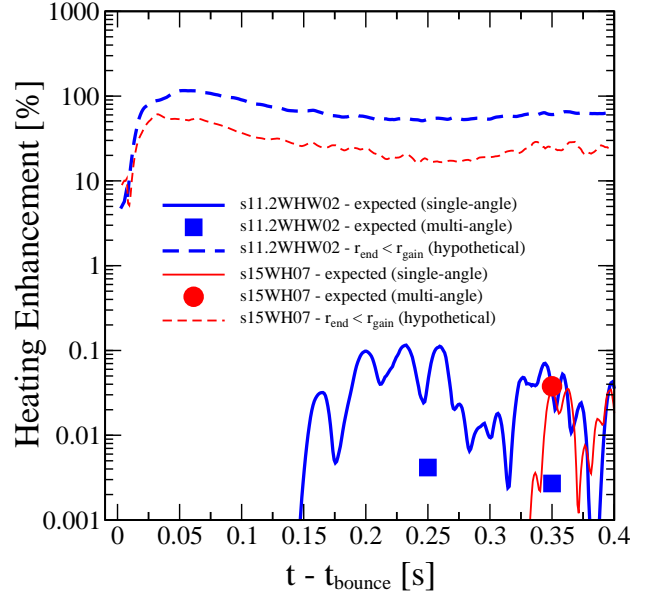}
\caption{Time evolution of the potential percentage increase in the
  heating rate, ${\cal {H}}_\mathrm{after}/{\cal{H}}_\mathrm{before} -
  1$, due to collective neutrino oscillation.  The dashed lines (thick
  for model s11.2WHW02 and thin for model s15WH07) assume the
  hypothetical case of complete conversion already below the gain
  radius, leading to an enhancement of $(20-100)\%$ depending on the
  progenitor and time after bounce. In our simulations, complete
  conversion does not occur before the gain radius. The more realistic
  estimate of the heating enhancement based on the oscillation
  calculations in Section \ref{collective} is much lower and shown in
  solid lines.  Before $t=150\,$ms and $330\,$ms the synchronization
  radius is outside the shock in model s11.2WHW02 and s15WH07,
  respectively. The points, blue squares for the s11.2WHW02
    model and red circles for the s15WH07 model, represent our
    estimate of the heating enhancement if the multi-angle survival
    probabilities are used.  At $150\,$ms for the s11.2WHW02 model and
    at $150$ and $250\,$ms for the s15WH07 model, no heating
    enhancement is seen, the conversion occurs completely outside the
    shock.}
\label{fig:enhancements}
\end{figure}

We employ the neutrino data from models s11.2WHW02 and s15WH07 and the
analytic approximations to the heating, Eqs.\,(\ref{eq:heat_before})
and (\ref{eq:heat_after}), to calculate the change in the heating rate
due to collective oscillations. The column number densities, $c_i$ and
$c_i^O$, going into the heating rates are angle-averaged values
obtained from simulation data. The results are depicted by
Fig.\,\ref{fig:enhancements}. Shown with dashed lines are the expected
heating enhancements in model s11.2WHW02 (thick lines) and model
s15WH07 (thin lines) in a \emph{hypothetical} scenario in which we
assume complete flavor conversion \emph{below} the gain radius
($P_\mathrm{ex} = 1$ everywhere). In this extreme case, the heating
would be enhanced by $\gtrsim$$60\%$ in model s11.2WHW02 and
$\gtrsim$$20\%$ in model s15WH07, this is similar to the
  configuration of Suwa~\emph{et al.}~\cite{suwa:11}. Note, however,
that our assumption that neutrinos of the accretion luminosity do not
undergo oscillations may be invalid in this hypothetical situation.

For a more realistic estimate of the heating enhancement, we use
$r_\mathrm{sync}$ as numerically computed for both models in
Section~\ref{sec:oscirad} and assume that above this radius
$P_\mathrm{ex} = 0.05$ and below this radius, $P_\mathrm{ex} = 0$.  As
a consequence, if $r_\mathrm{sync}$ is greater than the shock radius,
there is no heating enhancement.  The result of this is shown in
Fig.\,\ref{fig:enhancements} in thick-solid (thin-solid) lines for
model s11.2WHW02 (s15WH07). The predicted heating enhancement sets in
at much later times and is $\lesssim 0.1\%$ for both models. If one
instead assumed that $P_\mathrm{ex} = 1$ outside of $r_\mathrm{sync}$,
the enhancement would still be $\lesssim (2-3)\%$

When considering the predicted flavor conversion in our multi-angle
calculations discussed in Section~\ref{sec:maeffects}, we find, as
expected, that there is no further enhancement of the heating, but
rather that the enhancement is even more suppressed.  We show this in
Fig.~\ref{fig:enhancements}, where we denote by points the expected
heating enhancement using the multi-angle survival probabilities
presented in Fig.~\ref{fig:s11.2_and_s15-MA} at three postbounce times
for each progenitor. In the s11.2WHW02 model, the heating enhancement
is zero or reduced significantly.  The heating enhancement estimated
in the s15WH07 model, where we observed little change in the onset of
collective neutrino oscillations, is not as strongly affected. Hence,
we conclude that collective neutrino oscillations are very unlikely to
have a measurable effect on the neutrino heating and postbounce
dynamics in progenitors in and above the mass range considered in this
study.

\vspace{-0.4cm}
\section{Discussion and Conclusions} \label{sec:conclusions}

Almost eight decades after Baade \& Zwicky's stagesetting 1934 proposal
\cite{baade:34a,baade:34b} that a core-collapse supernova represents
the transition of an ordinary star to a neutron star, the details of
the mediating mechanism that converts gravitational energy of collapse
into energy of the core-collapse supernova explosion remain uncertain.

The neutrino mechanism, based on net heating by charged-current
neutrino absorption in the region just below the stalled shock,
appears to be the most viable candidate mechanism, requiring the least
special conditions (\emph{e.g.}, not requiring rapid rotation or strong
magnetic fields etc.) to succeed in exploding garden-variety Type-II
supernova progenitor stars. Yet, neutrino-driven explosions fail in 1D,
are marginal in 2D simulations, and modeling groups are now exploring
the neutrino mechanism's potentially boosted efficacy in
3D~\cite{nordhaus:10,fryerwarren:02}. While dimensionality may be
the key to successful explosions, it is also possible that current
1D/2D models are still missing some physics crucial to successful
explosions.

In this paper, we have considered new physics previously left out of
core-collapse supernova models: collective neutrino flavor
oscillations induced by neutrino-neutrino forward scattering in the
core-collapse supernova core. If the neutrino mass hierarchy is
inverted ($\dma<0$), collective oscillations will invariably lead to a
swap of $\nu_e$ and $\bar{\nu}_e$ spectra with the significantly
harder spectra of their heavy-lepton neutrino counterparts.  Assuming
a hypothetical scenario in which this swap is complete below the gain
region, we find that neutrino heating is enhanced by $(20-100)\%$ in
representative Type-II supernova progenitors
of $11.2$-$M_\odot$ and $15$-$M_\odot$. Such a significant boost of heating may
lead to strong, early explosions, breaking the strong feedback between
EOS, weak interactions, neutrino transport, and hydrodynamics that is
present in the postbounce phase of core-collapse supernovae and that
tends to absorb small variations in any of its
components\footnote{This behavior of strongly coupled astrophysical
systems is known as \emph{Mazurek's Law}~\cite{ott:09c}.}.
    
To study the viability of this scenario, we have performed collective
neutrino oscillation calculations in the single-angle and multi-angle
approximation on the basis of the neutrino radiation fields obtained
from 2D neutrino radiation-hydrodynamic core-collapse simulations
using $11.2$-$M_\odot$ and $15$-$M_\odot$ progenitors. From the
oscillation calculations we obtain the characteristic radii
$r_\mathrm{sync}$ and $r_\mathrm{end}$ at which $\sim$$5\%$ and
$\sim$$90\%$ of the flavor conversion have occurred, respectively. In
our simulations, these radii start hundreds of kilometers above the
typical shock radii in the early postbounce phase, but recede with
time as the neutrinospheres settle to smaller radii as the neutrino
luminosities decrease. The radius of onset of oscillations
($r_\mathrm{sync}$) reaches the average shock radius at
$(150-350)\,\mathrm{ms}$ and thereafter stays close to the latter,
while $r_\mathrm{end}$ is systematically $(150 - 200)\,\mathrm{km}$
outside the average shock radius. As a consequence, most of the flavor
conversion occurs outside the shock and can have no effect on the
postbounce heating and hydrodynamics. Those oscillations that take
place inside the shock occur close to the shock radius and, even when
taking large shock excursion driven by the SASI into account, enhance
the net heating by less than $\sim$$(2-3)\%$ in the most optimistic
case. These results strongly suggest that collective neutrino
oscillations are unlikely to have any qualitative or significant
quantitative impact on the postbounce evolution and the explosion
mechanism in the standard Type-II supernova progenitors
considered~here.

Our results also show that the characteristic oscillation radii assume
small values faster in cooler, less massive progenitors with lower
postbounce accretion rates and smaller neutrinosphere radii in the
postbounce phase. While oscillations are unlikely to boost the
pre-explosion neutrino heating in our $11.2$-$M_\odot$ and
$15$-$M_\odot$ models, the situation may be different in even
lower-mass progenitors with O-Ne cores or in O-Ne white-dwarf
progenitors of accretion induced collapse. The weak explosions already
obtained for such progenitors
\cite{kitaura:06,huedepohl:10,burrows:07c,dessart:06aic} could thus be
significantly enhanced.

As an aside, we have also shown that the multi-angle matter
suppression~\cite{EstebanPretel:2008ni, Chakraborty:2011nf,
  Chakraborty:2011gd} is somewhat sensitive to the choice of angular
emission spectrum and the matter density profile. This is expected to
affect the predictions of flavor evolution.

In deriving our results, we have made some approximations and
simplifications. Of these the most limiting is that we have assumed a
sharp neutrinosphere, common to all flavors and energies. This ignores
the interplay of oscillations and collisions, but a completely
self-consistent treatment would require a full collisional Boltzmann
calculation including oscillations, which must be left to future
work. Also, due to computational limitations, we have not carried out
multi-energy multi-angle oscillation calculations, which could be
improved upon in subsequent work.  The other significant limitation is
the assumption of axisymmetry in our supernova calculations. Future,
full 3D radiation-hydrodynamics simulations may lead to different
hydrodynamic postbounce evolutions and could yield different results.
Given that the understanding of core-collapse supernova physics and
collective neutrino oscillations is still in a state of rapid
development, there might be additional, yet unknown, effects that
could change our conclusions.

Finally, to summarize, collective neutrino oscillations are an
intriguing phenomenon. They have important important ramifications for
the core-collapse supernova neutrino signature, but, as we have shown
in this paper, they develop at too large radii to play a significant
role in the explosion mechanism.

\vspace{-0.4cm}
\acknowledgements We are indebted to E.~Livne and A.~Burrows for
kindly allowing us to use simulation results obtained with {\tt
  VULCAN/2D} in this work.  Furthermore, we acknowledge helpful
discussions with J.~Beacom, A.~Burrows, A.~Dighe, T.~Fischer,
E.~Livne, O.~Pejcha, G.~Raffelt, and T.~Thompson. BD
thanks A.~Mirizzi for a useful exchange of preliminary results. CDO wishes to thank 
M.~Kamionkowski for the inspiration to work on this subject and is
indebted to P.~Vogel for initial discussions and help with neutrino
oscillations. BD and CDO would like to thank the organizers of JIGSAW-2010 
at TIFR, Mumbai where this project was conceived.

This research is supported in part by the National Science Foundation
under grant nos.\ AST-0855535 and OCI-0905046 and by the Sherman
Fairchild Foundation.  EOC is supported in part by a post-graduate
fellowship from the Natural Sciences and Engineering Research Council
of Canada (NSERC).

Results presented in this article were obtained through calculations
on the Caltech compute cluster ``Zwicky'' (NSF MRI award
No.~PHY-0960291), on the Louisiana Optical Network Initiative
supercomputing systems under allocation loni\_numrel06, on the NSF
TeraGrid under award No.~TG-PHY100033, and on resources of the
National Energy Research Scientific Computing Center, which is
supported by the Office of Science of the U.S. Department of Energy
under Contract No. DE-AC02-05CH11231.



\end{document}